\documentclass[ALICE,manyauthors]{cernphprep}

\usepackage[comma,square,numbers,sort&compress]{natbib}
\usepackage{hyperref}
\usepackage[utf8]{inputenc}

\pdfoutput=1

\newcommand{\pT}{\ensuremath{p_{\mbox{\tiny T}}}\xspace}
\newcommand{\GeVc}{\ensuremath{\mbox{GeV}/c}\xspace}

\begin{document}%

\begin{titlepage}
\PHyear{2019}
\PHnumber{020}      
\PHdate{8 February}  
%

\title{Calibration of the photon spectrometer PHOS of the ALICE experiment}
\ShortTitle{Calibration of the photon spectrometer PHOS of the ALICE experiment}   

\Collaboration{ALICE Collaboration\thanks{See Appendix~\ref{app:collab} for the list of collaboration members}}
\ShortAuthor{ALICE Collaboration} 

\begin{abstract}
  The procedure for the energy calibration of the high granularity electromagnetic calorimeter PHOS of the ALICE experiment is presented.
  The methods used to perform the relative gain calibration, to evaluate the geometrical alignment and the corresponding correction of the absolute energy scale, 
  to obtain the nonlinearity correction coefficients and
finally, to calculate the time-dependent calibration corrections, are 
discussed and illustrated by the PHOS
  performance in proton-proton (pp) collisions at $\sqrt{s}=13$~TeV. 
After applying all corrections, the achieved mass resolutions for $\pi^0$ and $\eta$ mesons for $\pT > 1.7$~\GeVc are $\sigma_m^{\pi^0} = 4.56 \pm 0.03$~MeV/$c^2$ and $\sigma_m^{\eta} = 15.3 \pm 1.0$~MeV/$c^2$, respectively.
\end{abstract}
\end{titlepage}
\setcounter{page}{2}

\section{Introduction}

The ALICE experiment \cite{Aamodt:2008zz} is one of the four major experiments at the Large Hadron Collider (LHC) at CERN. 
Its primary goal is the study of the properties of the hot and dense quark--gluon matter created in ultrarelativistic heavy-ion collisions. This dictates the unique features of the ALICE detector design: ability to register and identify both soft particles, reflecting collective behavior of
the hot matter, and hard penetrating probes, i.e. jets, direct photons, etc., carrying information about the inner, hottest part of the created fireball. 
The ALICE experiment incorporates detectors based on a number of 
particle identification techniques. The tracking system is able to detect
and identify relatively soft charged particles with transverse momenta $\pT>50-100$~MeV/$c$ and process high-multiplicity events. 
ALICE includes an electromagnetic calorimeter system: the PHOton
Spectrometer (PHOS) \cite{Dellacasa:1999kd,Aamodt:2008zz} and the Electromagnetic
Calorimeter (EMCal) \cite{Cortese:2008zza} with the Di-Jet Calorimeter (DCal) \cite{Allen:2010stl}. 
The PHOS calorimeter is designed to measure spectra, collective flow and correlations of thermal and prompt direct photons, and of neutral mesons via their decay into photon pairs. 
This requires high granularity as well as excellent energy and position resolution. 
The electromagnetic calorimeter EMCal/DCal is used for the measurement of electrons from heavy flavour decays and the electromagnetic component of jets, spectra and correlations of isolated direct photons and spectra of neutral mesons. This requires a large acceptance but less strict requirements on the energy and position resolution. 
In this paper, the methods used for the calibration of the PHOS detector during the LHC data taking campaigns of 2009$-$2013 (Run~1) and 2015$-$2017 (Run~2) are described and results of the calibration are presented.


The procedure for electromagnetic calorimeter calibration, developed by
high-energy experiments, depends on physics objectives, detector resolution, beam
availability and hardware implementation of the calorimeters and their
front-end electronics.
The four LHC experiments use different approaches:
the electromagnetic calorimeter (ECAL) of the LHCb experiment \cite{LHCB:2000ab}
was pre-calibrated with an energy flow method, 
requiring the transverse energy distribution over the calorimeter to be a 
smooth function of the coordinates. A final detailed calibration was carried out using the $\pi^0$ peak, using
invariant mass distributions and the minimization of event-by-event variables \cite{Pereima:2017dvq, Belyaev:2011zz}.
The electromagnetic calorimeter (ECAL) of the CMS experiment \cite{CMS:1997ema} was pre-calibrated with laboratory measurements 
of crystal light yield, and the gain and quantum efficiency of the photodetectors. These were followed by beam tests with high-energy electrons and cosmic-ray muons. The absolute calibration was determined by using the $Z$-boson mass and channel-by-channel relative calibration. The relative calibration involved the measurement of transverse energy and the use of $\varphi$-symmetry, the $\pi^0$ and $\eta$ meson invariant mass fit, and a comparison of the energy measured in the ECAL to the track momentum measured in the silicon tracker for isolated electrons from $W^-$ and $Z$-boson decays \cite{Fasanella:2017gqq, Chatrchyan:2013dga}.
The longitudinally segmented liquid-argon calorimeter of the ATLAS experiment \cite{ATLAS:1996ab} was calibrated by using a multivariate algorithm to simulate the $e/\gamma$ response \cite{Aad:2014nim}. The absolute energy scale was calibrated by using electrons from a large
sample of $Z \rightarrow e^+e^-$ decays and validated with $J/\psi \rightarrow e^+e^-$ decays.

The energy calibration of PHOS includes four mutually dependent aspects: relative gain calibration, absolute energy calibration, nonlinearity correction, and time-dependent calibration correction. 
The PHOS detector will be briefly described in section \ref{Setup}. 
The relative gain calibration is presented in section \ref{sec:Relative}, 
including the pre-calibration using the LED monitoring system and
the calibration using the $\pi^0$ peak position which are described in
sections \ref{sec:LED} and \ref{sec:Pi0}, respectively. 

Fixing the absolute energy calibration of a calorimeter using the $\pi^0$ mass peak suffers from systematic uncertainties due to the geometrical alignment of the calorimeter and the energy scale. Because of that the absolute energy calibration is validated using the electron $E/p$ ratio, as described in section \ref{sec:Electron}, and the detector geometrical alignment is checked as described in section \ref{sec:Align}. 
The estimation of the nonlinearity correction is described in section \ref{sec:Nonlin} and the calculation of the time-dependent energy calibration correction is discussed in section \ref{sec:Run-by-run}. The final calibration results are presented in section \ref{sec:Results}. 

\section{Setup}
\label{Setup}

\begin{figure}[ht]

  \includegraphics[width=\hsize]{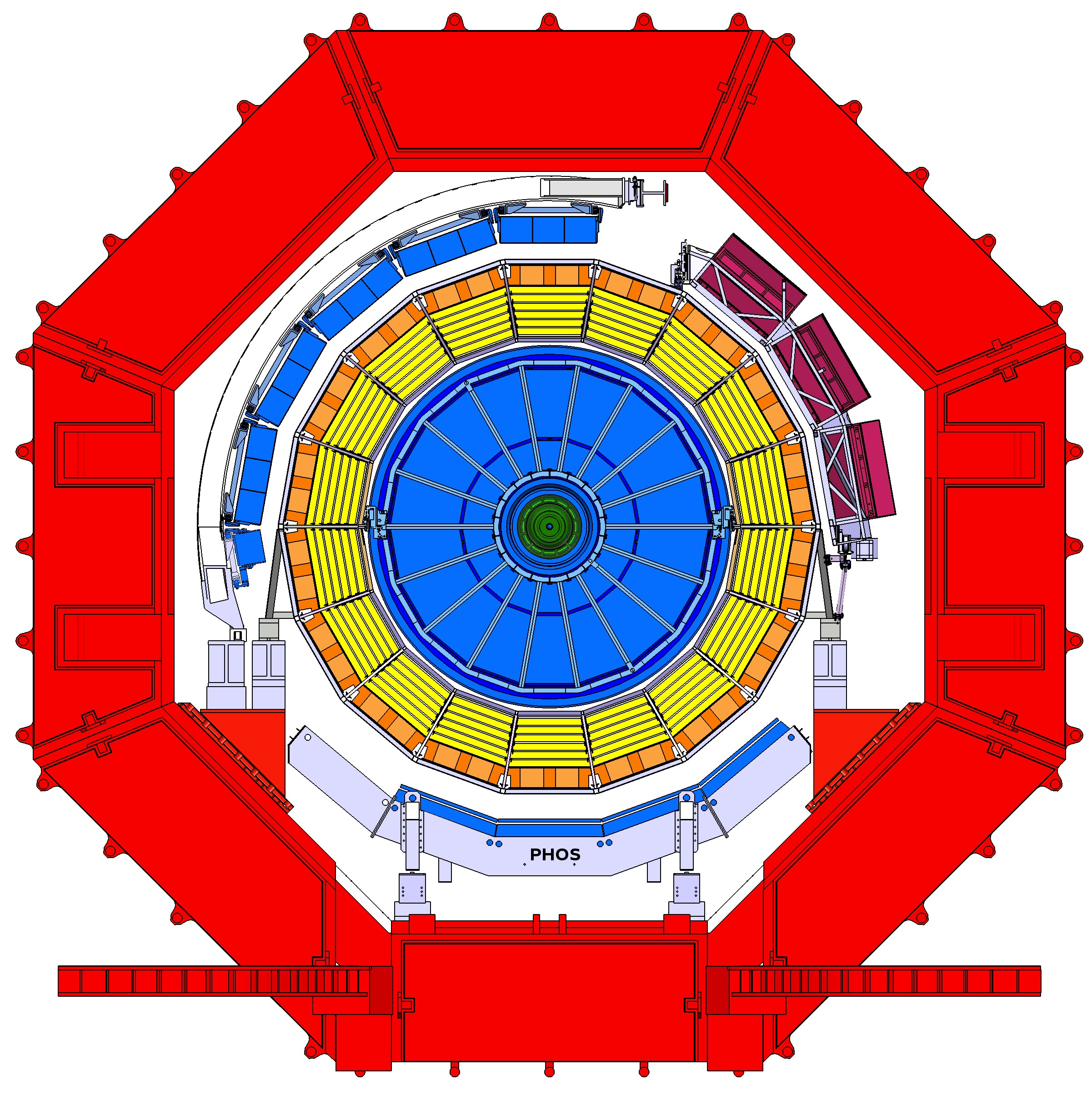}
  \caption{[Color online] ALICE cross-sectional view in Run 2, PHOS modules are located at the bottom of the setup.}
  \label{fig:SetupALICE}
\end{figure}

The PHOS is a single arm, high-resolution electromagnetic calorimeter which detects and identifies
photons and electrons in a wide \pT range from $\sim 100$ MeV/$c$ to $\sim 100$ GeV/$c$
at mid-rapidity and, additionally, provides a trigger in case of a large energy deposition by an energetic particle.
%
The main parameters of the detector are summarized in Tab.~\ref{phos:tab-PHOSparam}. PHOS is located inside the solenoid magnet providing a 0.5 T magnetic field. The TRD and TOF detectors are designed to have windows in front of the PHOS modules to reduce the material budget in front of the PHOS down to 0.2 $X_{0}$ \cite{Acharya:2017lco}. The PHOS is subdivided into four
independent units, named modules, positioned at the
bottom of the ALICE detector at a radial distance of 460~cm from the
interaction point (IP) to the front surface of crystals as shown in Fig.\ \ref{fig:SetupALICE}. It covers approximately a quarter of a unit
in pseudo-rapidity, $|\eta| \le 0.125$, and 70$^\circ$ in
azimuthal angle. Its total active area is 6~m$^2$.

\begin{table}[h!]
  \caption{General parameters of the PHOS detector}\label{phos:tab-PHOSparam}
  \centering
  \begin{tabular}{l l}
  \hline
  \hline
  Coverage in pseudo-rapidity \qquad & $-0.125 \le \eta \le 0.125$ \\
  Coverage in azimuthal angle & $\Delta\varphi =$70$^\circ$ \\
  Distance to interaction point & 460~cm \\
  Modularity & Three modules with 3584  and one with 1792 crystals\\
  \hline
  Material & Lead-tungstate (PbWO$_4$) crystals \\
  Crystal dimensions & $22\times 22 \times 180$~mm$^3$ \\
  Depth in radiation length & $20$ $X_0$ \\
  Number of crystals & 12\,544 \\
  Total area    & 6.0 m$^2$ \\
  Operating temperature & $-25^\circ$ C \\
  \hline
\hline
\end{tabular}
\end{table}

\begin{figure}[hb]
    \includegraphics[width=0.505\hsize]{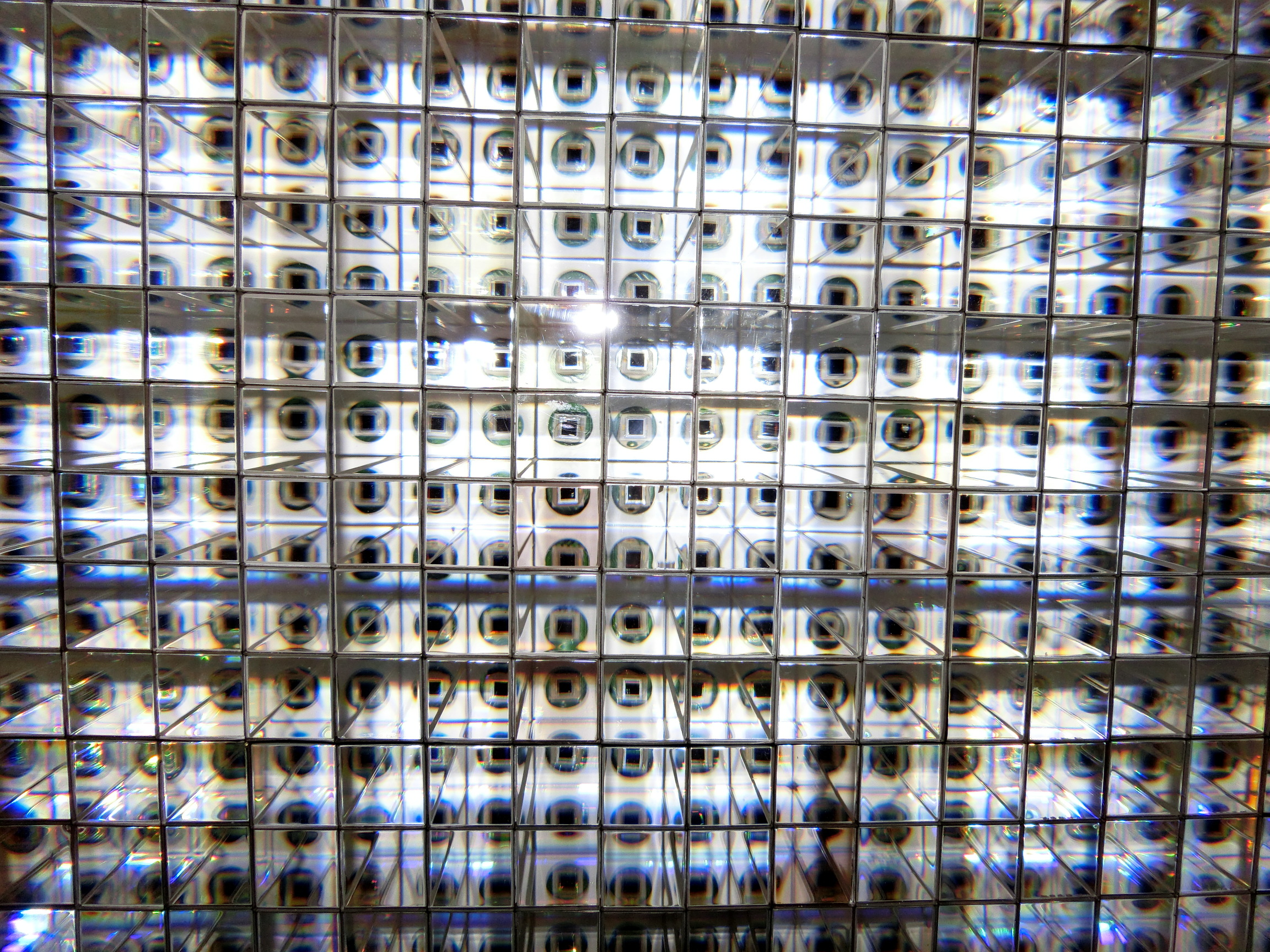} 
    \hfill
    \includegraphics[width=0.48\hsize]{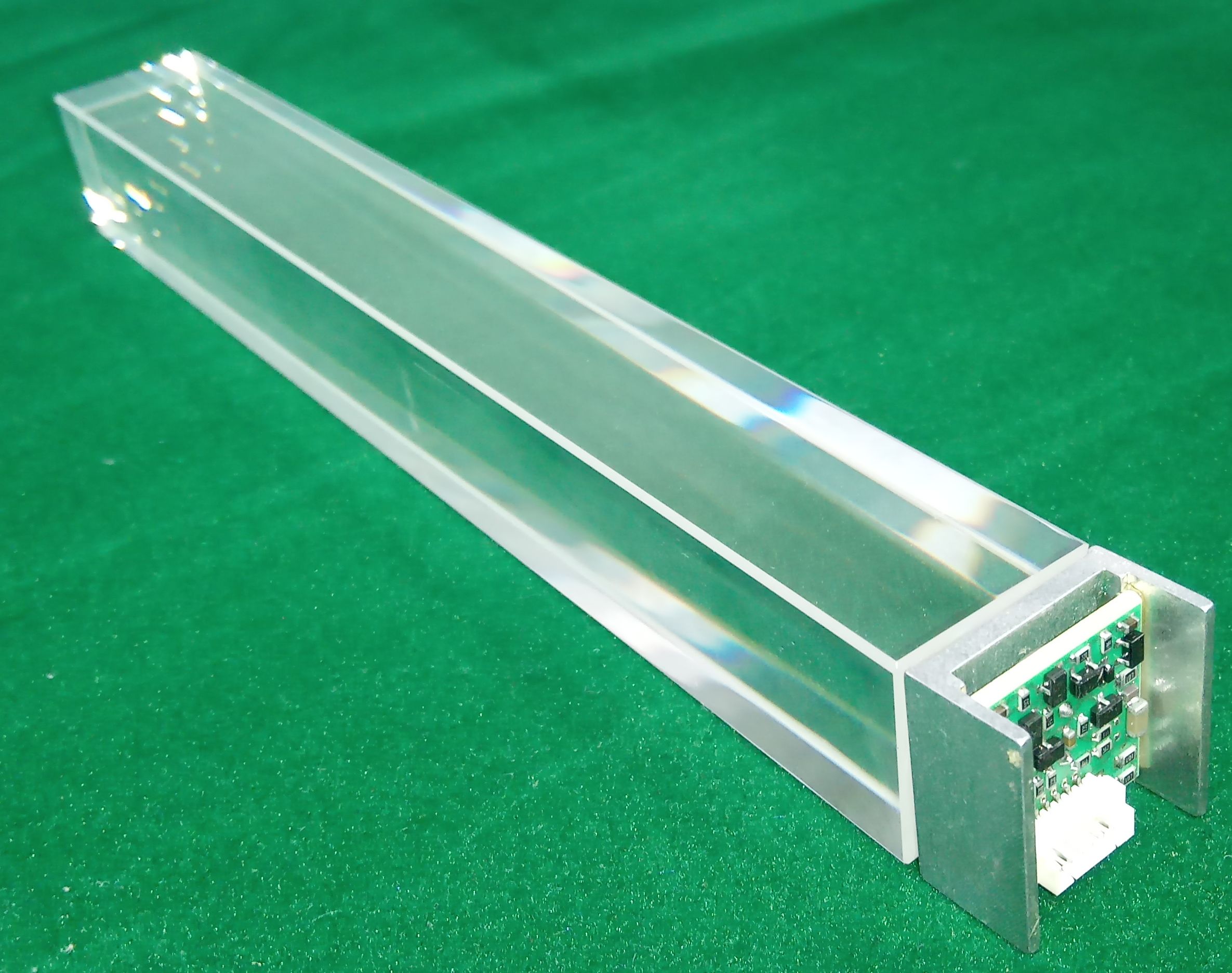} 
  \caption{[Color online]
    Left: Part of a cell matrix of one module; Right: A detector element comprising a {\rm
PbWO}$_4$ crystal, APD photodetector and preamplifier.}
  \label{fig:Setup}
\end{figure}

Three PHOS modules are segmented into 3584 detection elements (cells) arranged
in 56 rows of 64 elements each, while the fourth module has 56 rows of 32 elements.
A part of a cell matrix is shown in Fig.\ \ref{fig:Setup}, left.
The PHOS modules are numbered counterclockwise in Fig.~\ref{fig:SetupALICE} \cite{Aamodt:2008zz}.
Each detection element comprises a $22\times 22\times 180$~mm$^3$ lead-tungstate crystal, {\rm
PbWO}$_4$ \cite{Ippolitov:2005wi}, coupled to a $5\times 5$~mm$^2$ Avalanche
PhotoDiode (APD Hamamatsu S8664-55) whose signal is processed by a low-noise
preamplifier. 
The APD and the preamplifier are
integrated in a common body glued onto the end face of the crystal
with optically transparent glue with a high refractive index, see Fig.\ \ref{fig:Setup}, right. 
The PbWO$_4$ 
was chosen as an active
medium because it is a dense, fast and
relatively radiation-hard scintillating crystal. 
Its radiation length is only 0.89~cm and its Moli\`ere radius is 2.0~cm. It has a broad emission
spectrum with bands around 420 and 550~{\rm nm}\ \cite{Ippolitov:2005wi}.
%

The light yield of PbWO$_4$ crystals is relatively low and strongly depends
on temperature (temperature coefficient of $-2\%/^\circ$C). 
In order to increase the light yield by about a factor 3 compared to standard conditions, the PHOS
crystals are operated at a temperature of $-25^\circ$C. 
The energy resolution of a PHOS prototype measured under these conditions in beam tests \cite{PHOS-beam-test}
is described by a parametrization as follows
\begin{equation}
  \frac{\sigma_{E}}{E} = \sqrt{\left(\frac{a}{E}\right)^{2} + \left(\frac{b}{\sqrt{E}}\right)^{2} + c^{2}}
  \label{Eres}
\end{equation}
where $a=0.013$~GeV, $b=0.0358$~GeV$^{1/2}$ and $c=0.0112$. 
The temperature of the PbWO$_4$ crystals is stabilized with a precision of $~0.3^\circ$C.
Temperature monitoring is based on resistive temperature sensors of
thickness 30$-$50 $\mathrm \mu$m inserted in the gap between the
crystals. For the purpose of temperature stabilization, a PHOS
module is subdivided by thermo-insulation into ``cold'' and ``warm''
volumes. Strip units, comprising two rows of eight detection elements, are mounted
onto the main mechanical assembly points in a module. The crystal strips are located in the
cold volume, whereas the readout electronics are located outside, in the warm volume.
The APDs belonging to one strip unit, and their associated
preamplifiers, provide $2\times 8$ analog signals to a T-shaped
connector which passes the signals from the cold zone to the front-end
and trigger electronics located in the warm zone.  All six sides of
the cold volume are equipped with cooling panels. The heat is
removed by a liquid coolant (perfluorohexane, C$_6$F$_{14}$)
circulating through the pipes on the inner panel surfaces. Moisture
condensation is prevented by making airtight cold and warm
volumes ventilated with nitrogen.

Every channel in the PHOS detector is monitored with an LED system that
is driven by stable current injectors~\cite{Bogolyubsky:2012zz}. The system consists of LED matrices for
each PHOS module, having one LED per PHOS cell with controlled light
amplitude and flashing rate.

The PHOS electronic chain includes energy digitization and trigger
logic for generating trigger inputs to the zero (L0) and first (L1)
levels of the ALICE Central Trigger Processor
(CTP)~\cite{ALICEtriggerTDR}. In order to cover the required large
dynamic range from 10 MeV to 100 GeV, each energy shaper channel
provides two outputs with low and high amplification, digitized in
separate ADCs. The upper limit of the dynamic ranges in high- and
low-gain channels are 5~GeV and 80~GeV, with the ratio of these amplifications varying slightly from
channel to channel with a mean of approximately 16.8.
The gain of each APD can be set individually, by adjusting the bias voltage through the voltage distribution and control system. 
To equalize the energy response of all cells, the bias voltage of each APD
can be set to a precision of 0.2~V, which corresponds to a $\sim 0.5$\% gain variation (see Fig.~\ref{fig:LEDvsHV}, left for more details).
The timing information is derived from
an offline pulse-shape analysis.

\section{Energy calibration procedure}
\label{sec:Relative}

Photons and electrons hitting an electromagnetic calorimeter produce electromagnetic
showers with a transverse profile determined by the Moli\`ere radius of the calorimeter material. When the transverse cell size of the calorimeter is
comparable with the Moli\`ere radius, such as in PHOS, the electromagnetic shower is
developed in several adjacent cells around the impact 
point. The group of cells with common edges, containing the electromagnetic shower
generated by a photon, is referred to as a cluster (see sec. 4.5.2 of \cite{Dellacasa:1999kd}). The sum of
energies deposited by the shower in each cell of the cluster, 
is the measured photon energy \cite{Alessandro:2006yt}. 
With the PHOS granularity, the energy deposited in the central
cell of the cluster is about 80\% of the total cluster energy.


The amplitude of the signals measured in the cells of the cluster is proportional to the deposited energy in the cells. A set of calibration procedures is necessary to convert these data to an appropriate energy scale. 
Relative energy calibration means equalization of the response of all
channels to the same energy deposition. 
In the case of PHOS, calibration at the hardware level via adjusting
the APD bias voltage is complemented by refinement of the calibration
parameters in an offline analysis.
In order to ensure the uniformity of trigger response over the PHOS acceptance, the amplification in all channels was adjusted to make the trigger efficiency response turn-on curve as sharp as possible. 
This adjustment was performed once during the PHOS commissioning in LHC Run 1 and just before the start of the LHC Run 2 data taking period. The final calibration is done in an offline analysis described hereafter in detail.
In order to disentangle calibration effects from effects related to cluster overlaps in the high occupancy environment of heavy-ion collisions, the calibration is performed in low occupancy events provided by pp collisions.
 
At first, two approaches were tested: calibration using the
Minimum-Ionizing Particle (MIP) peak and equalization of mean energies
in each channel.
The minimum ionization signal of charged particles in the PHOS detector has a most probable value of about 250 MeV which is close to the lower end of the dynamic range. The calibration based on the MIP peak has a poor accuracy because of several effects such as relatively low number of counts of charged particles per cell, low signal-to-background ratio of the MIP signal in the PHOS energy spectrum, a wide spread of incident angles of charged particles which lead to the MIP energy variation.
The second method, based on the mean energy equalization, had a poor convergence and large uncertainties on the calibration parameters. Without pre-calibration using the APD gain adjustment, the mean energy strongly depends on the range of averaging which, in turn, depends on the initial calibration.
Nevertheless, this method was used to provide a reasonable calibration for the first measurement of neutral meson spectra in 
2010~\cite{Abelev:2012cn}, when the accumulated number of counts was not sufficient for more precise methods. Later, a more precise calibration based on the $\pi^0$ peak equalization described below was deployed in all subsequent papers \cite{Abelev:2014ypa,Acharya:2017tlv,Acharya:2018hzf,Adam:2015lda,Acharya:2018bdy}.

Our final strategy of the PHOS relative calibration is based on APD gain equalization as a pre-calibration (see
section \ref{sec:LED}) and the $\pi^0$ peak adjustment as a final 
step (see section \ref{sec:Pi0}).

\subsection{Gain ratio calibration}
\label{sec:HLratio}

\begin{figure}[ht]
  \centering
  \includegraphics[width=0.6\hsize]{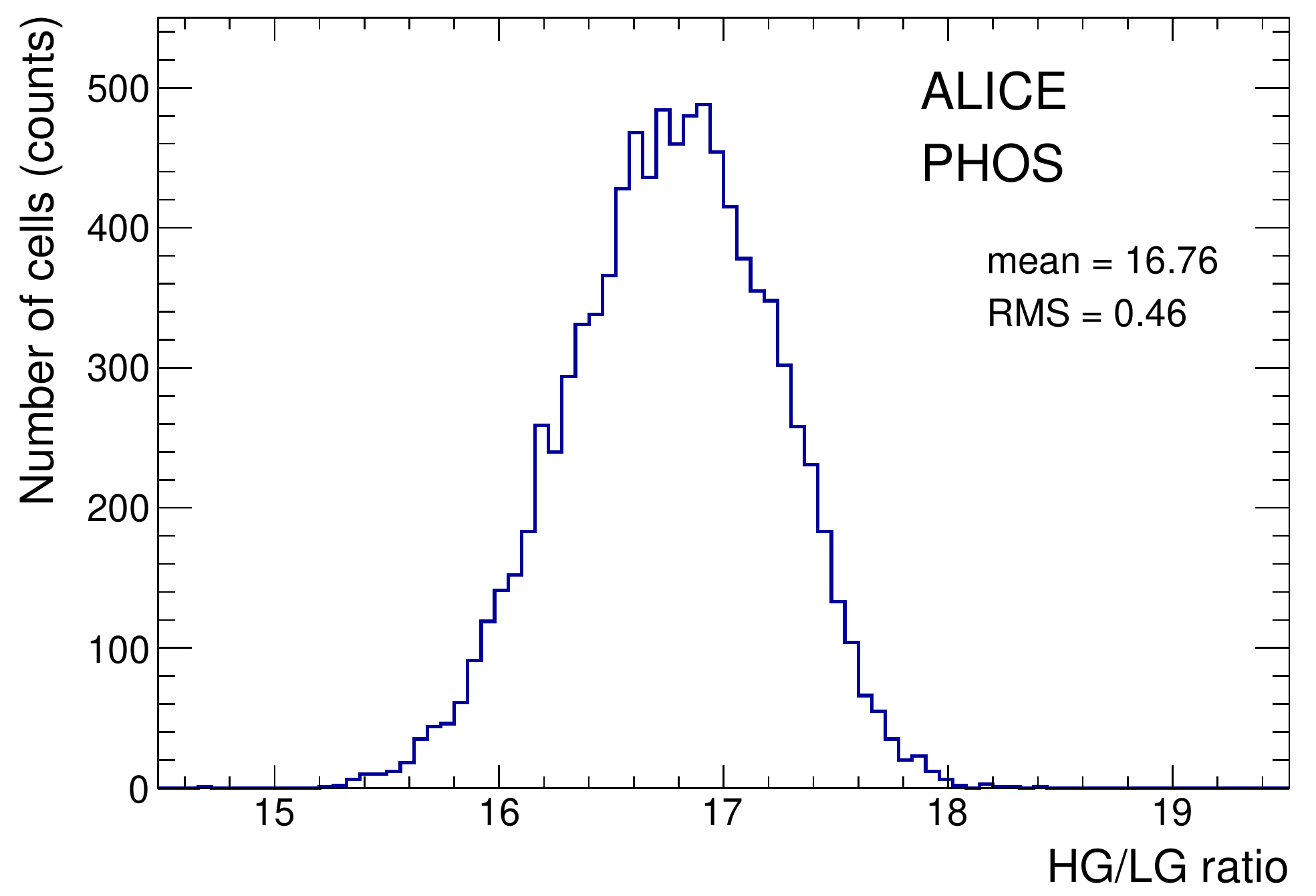}
  \caption{The ratio of high-to-low gains, for all cells.}
  \label{fig:HLratio}
\end{figure}

The LED monitoring system, with its capability to emit signals at high rate
and with variable amplitudes covering the whole dynamic range of the PHOS, was used to measure the high-to-low gain ratio. The gain ratio distribution for all active PHOS cells is shown in Fig. \ref{fig:HLratio} and spans from 15 to 18
with an average of about 16.8. The gain ratio is used for high energy amplitudes exceeding the high-gain dynamic range. In this case,
the energy is the product of the high-gain calibration parameter and the high-to-low gain ratio. The high-to-low gain ratio is stable and does not need to be frequently measured and updated.

The ratio of high-to-low gain is defined by the electronics components of the amplifiers and may vary from channel to channel. Therefore it is considered as one of the calibration parameters to be determined. The calibration methods discussed in the section~\ref{sec:Pi0} of this paper are based on data collected with beam, and ensure a good calibration of high-gain channels within the high-gain dynamic range, $E<5$~GeV. Low-gain channels can hardly be calibrated with the $\pi^0$ peak adjustment method described in section~\ref{sec:Pi0}, because of the limited statistics of high-energy clusters. Therefore the ratio of high-to-low gain has to be measured independently using signals of amplitudes which are detected simultaneously in both high- and low-gain channels.

\subsection{Photodetector gain equalization}
\label{sec:LED}

\begin{figure}[ht]
  \centering
  \includegraphics[width=0.6\hsize]{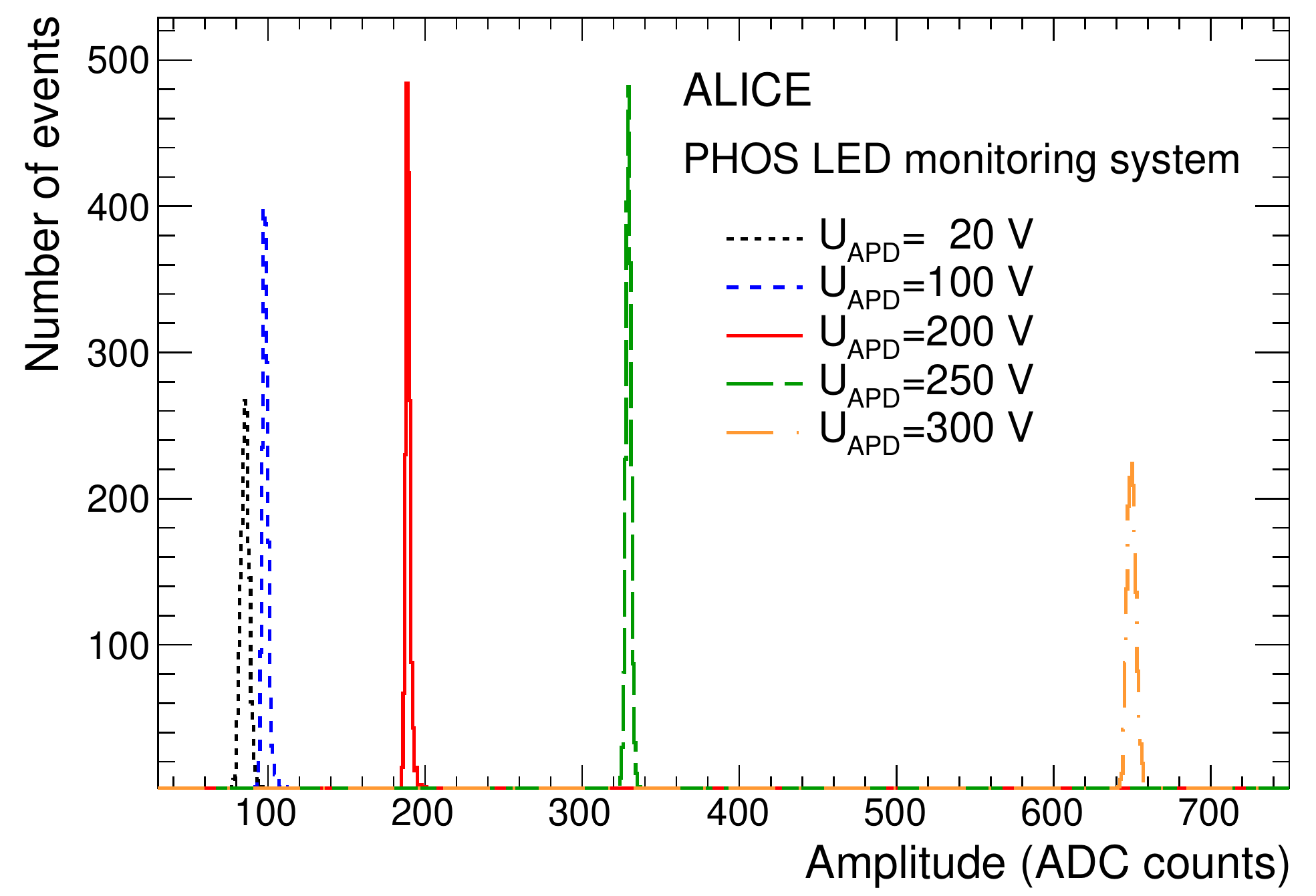}
  \caption{[Color online] The amplitude of the LED peak for different APD bias voltages, for one
example channel.}
  \label{fig:LEDpeaks}
\end{figure}

Each APD has a particular gain dependence on bias voltage. At low bias voltages, the APD gain is assumed to be unity. The APD gain is calculated as the ratio of the measured amplitude at a
given voltage to a reference amplitude at 20~V where the 
dark current in the APD is negligible. 
The dependence of the APD gain on the bias voltage was measured using 
the PHOS LED monitoring system, whose programmable light output was shown to be very stable
over several hours, a period far longer than necessary for gain measurements. 
The amplitude distribution from the LED flash is
measured at several values of APD bias voltage in the range from 20 to
395~V. Figure~\ref{fig:LEDpeaks}~shows the LED amplitude 
for different voltages for one example channel.

\begin{figure}[h]
  \centering
  \includegraphics[width=0.49\hsize]{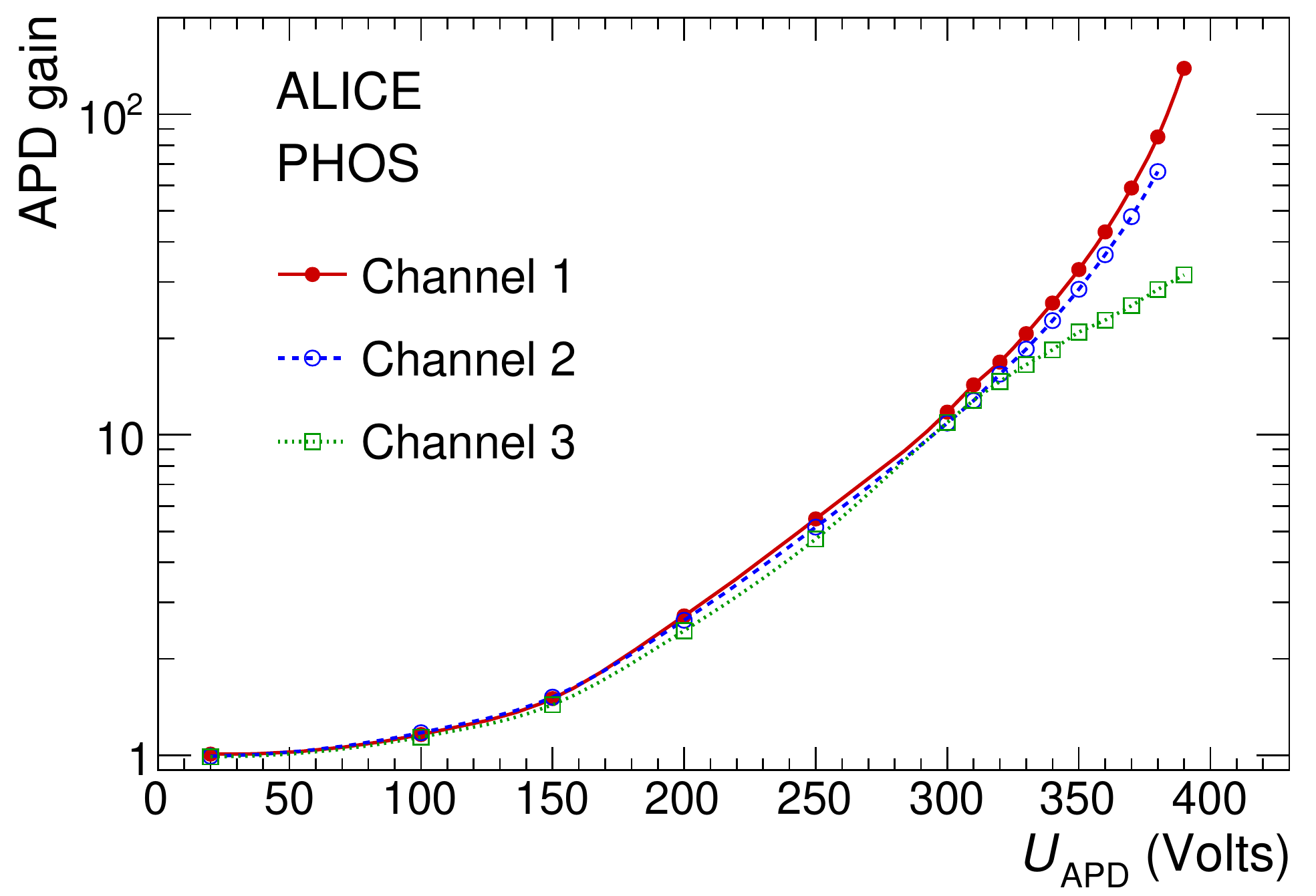}
  \hfil
  \includegraphics[width=0.49\hsize]{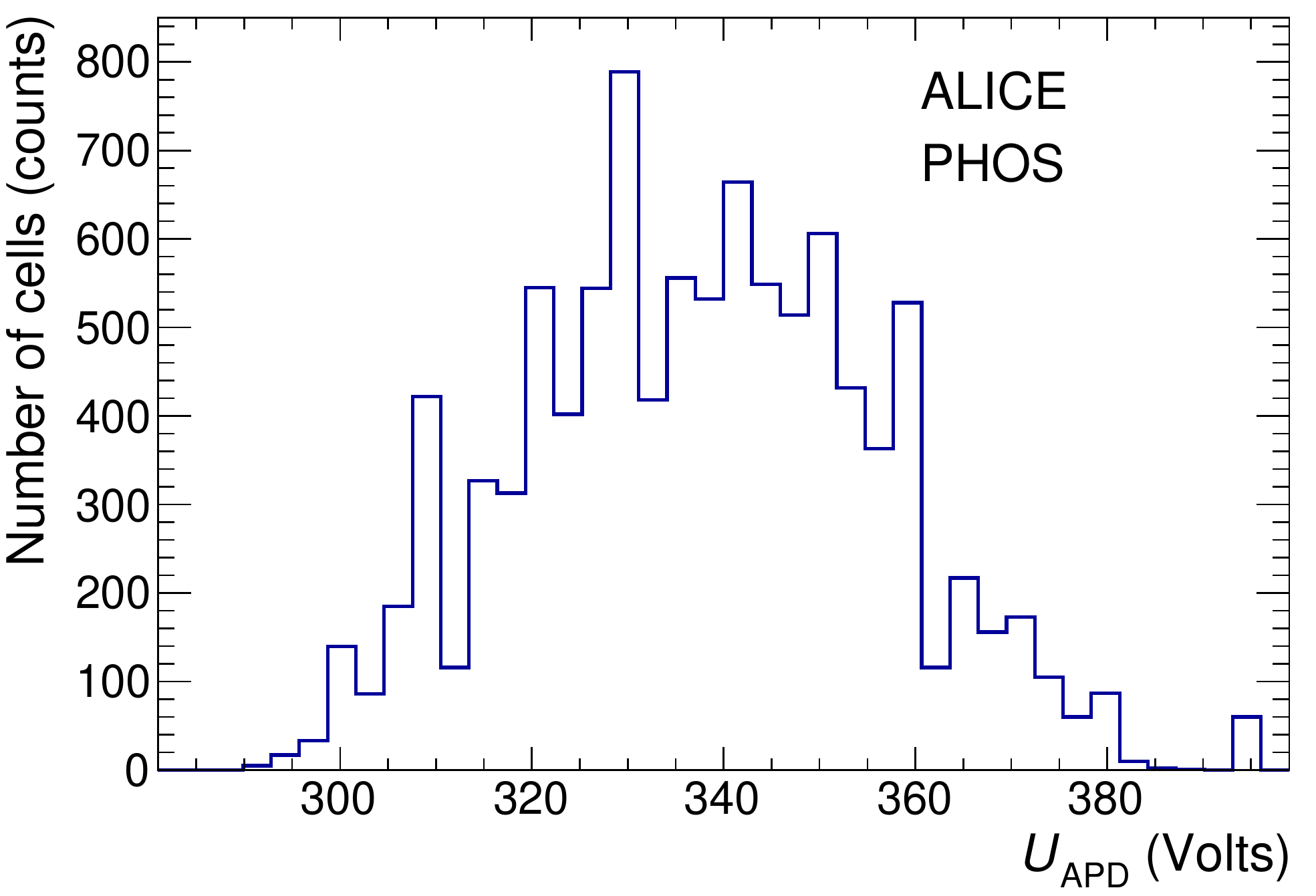}
  \caption{[Color online] Left: The dependence of the APD gain on applied bias voltage, for three different channels. Typical and two extreme cases are presented. Right: The distribution of the APD bias voltages, for all PHOS cells,
for an APD gain of 29.}
  \label{fig:LEDvsHV}
\end{figure}

Figure~\ref{fig:LEDvsHV} (left)
shows the gain-voltage dependence for three channels 
illustrating the spread of the gains at a given voltage.
An APD gain of 29 was set for all channels in order to provide the
designed dynamic range of the energy measurement in PHOS.
The bias voltages are required to cover a range 
from from 290 to 395~V,
as shown in Fig.~\ref{fig:LEDvsHV} (right).

\begin{figure}[t]
  \centering
  \includegraphics[width=0.5\hsize]{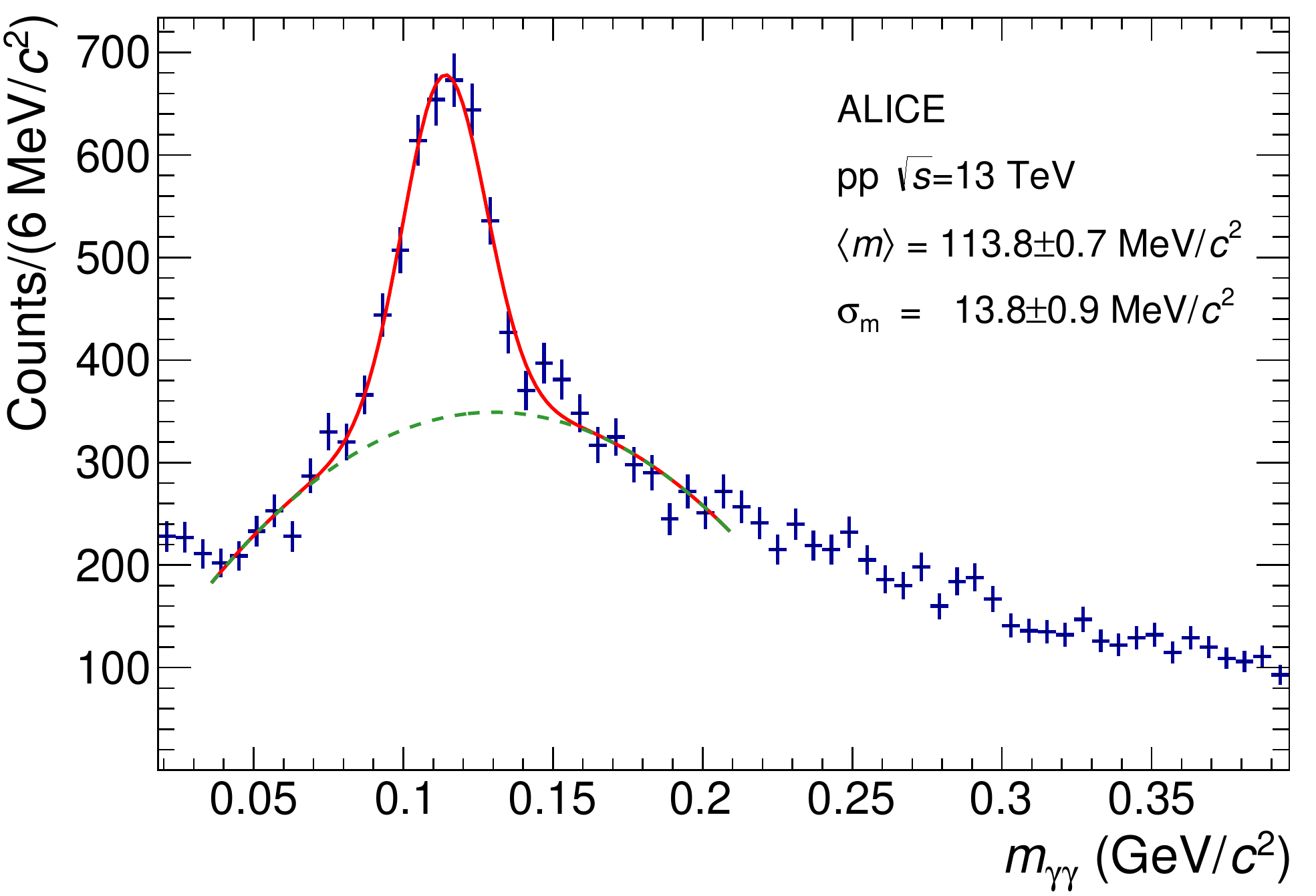}
  \caption{[Color online] Invariant mass distribution of cluster pairs after APD gain
    equalization in pp collisions at $\sqrt{s}=13$ TeV for $\pT>1.7~\GeVc$. The red curve is a fit of the spectrum
    with the sum of a Gaussian and a second-order polynomial function. The green dashed line is the background contribution only.}
  \label{fig:pi0PeakAPDgains}
\end{figure}
After the equalization of the APD gains, the calibration needs to be further refined to take into account the specific light yield of the different crystals. 
However, the spread of light yields of the different PbWO4 crystals is  about 12\% \cite{Ippolitov:2005wi}, which is relatively small compared to the initial pre-calibration, and has been neglected. 
The APD gain equalization
can thus be considered as a first step towards the energy calibration based
on physics signals from collision events such as the $\pi^0$ peak. 

The invariant mass of photon pairs is constructed as follows:
\begin{equation}
  m_{\gamma\gamma}=\sqrt{2E_{\gamma,1} E_{\gamma,2}(1-\cos \theta_{12})}, 
  \label{eq:mgg}
\end{equation}
where $E_{\gamma,i}$ is the energy of the reconstructed photon $i$, and 
$\theta_{12}$ is the opening angle between the two photons. 
The invariant mass distribution of cluster pairs detected in PHOS was measured in pp collisions, at $\sqrt{s}=13$ TeV, with a cut on the cluster pair transverse momentum of $\pT>1.7~\GeVc$.
Fig.~\ref{fig:pi0PeakAPDgains} shows the invariant mass distribution after APD gain equalization. The choice of the low-\pT cut
is driven by maximizing the signal-to-background ratio and minimizing the energy nonlinearity effects which will be
discussed in Section~\ref{sec:Nonlin}.
A clear $\pi^0$ peak above the combinatorial
background is observed. 
The invariant mass distribution is fitted in the range $35-210$ MeV/$c^2$ with the sum of a Gaussian and a second order polynomial. 
The extracted $\pi^0$ peak position $\langle m \rangle \approx
113.8\pm 0.6$~MeV/$c^2$ is $\sim 15$\% lower than the PDG value~\cite{Tanabashi:2018oca} and its width $\sigma_m \approx 13.8\pm 0.9$~MeV/$c^2$ is approximately 
3 times larger than the expected resolution of 5.5 MeV/$c^2$ for an
ideally calibrated PHOS as described in GEANT-based Monte Carlo simulations~\cite{Alessandro:2006yt}.
However, these values are an acceptable starting point for the final relative PHOS calibration based on $\pi^0$ peak
equalization described in the following section.

\subsection{Calibration using the $\pi^{0}$ peak position}
\label{sec:Pi0}

The calibration procedure calculates the calibration coefficient $\alpha_{i}$ relating
the energy deposition $E_{\rm dep}$ and the measured response amplitude, $A$, with
$E_{\rm dep}=\alpha_i\cdot A$, for each detector channel. 
To find the coefficients, the di-photon invariant mass distribution is constructed, see
Eq. (\ref{eq:mgg}). One of the two photons must directly hit the detector channel under consideration. The second photon can be anywhere in PHOS.

The invariant mass distribution shows a peak corresponding to the $\pi^0$ meson at $m_i$ with some mass shift due to miscalibration. A correction to the calibration coefficient, which relates the measured amplitude $A$ and corrected
energy $E_{\rm corr}$ as $E_{\rm corr}=\alpha_i\cdot c_i \cdot A$, 
is defined by the following equation: 
\begin{equation}
  c_{i}=\left ( \frac{m_{\pi^0}}{m_i} \right )^{n},
  \label{ccoef}
\end{equation}
where $m_{\pi^0}$ is the true neutral pion mass and $n>0$ is a parameter that has to be optimized. 
The procedure is iteratively applied, with $\alpha_i$ obtained at iteration $j$ being updated to 
$\alpha_{i}^{j+1} = \alpha_i^{j} \cdot c_i$, 
until no further improvement of a calibration is found. 
If we assume that the calibration coefficients (for all cells where partner photons are registered) fluctuate around some mean value, and therefore their energies are correct on average, then the shift of the peak position can be attributed to the miscalibration of the current cell. From Eq. (\ref{eq:mgg}) $E_{1,\gamma}=m/m_{\mathrm i}^2/(2E_{2,\gamma}(1-\cos \theta_{12})$,  the correction coefficient for the current cell $i$ is $c_{\mathrm i}=E_{\mathrm correct} /E_{\mathrm i} = m_{\mathrm PDG}^2/m_{\mathrm i}^2$ and  one can expect that the most appropriate power is $n=2$. However, this assumption is not completely true. To illustrate this, the procedure is applied to a toy model implementing several values of $n$ as described in the next section.

\subsubsection{Optimization of the calibration procedure with a toy model}
The toy model describes the influence of the simultaneous calibration of different cells of a calorimeter. In a real calorimeter a photon cluster includes a cell with a dominant energy deposition plus a few additional cells.
The simplified model assumes that the entire photon energy is deposited in one cell of a calorimeter.   
In the model, the calorimeter covers a pseudorapidity $|\eta|<1$ and full azimuthal angle with a granularity of $100\times 100$ cells in the $\varphi$ and $\eta$ directions. Each cell has an independent calibration coefficient which initially is randomly assigned according to a Gaussian distribution with 
mean 1 and a width of 20\%. 

The particle generator is tuned to produce neutral pions with a flat rapidity and azimuthal distribution and a realistic $\pT$ spectrum as measured in pp collisions at $\sqrt{s}=7$ TeV \cite{Abelev:2012cn}. 
The generated $\pi^0$ mesons are forced to decay into photon pairs. The photon energies are smeared according to Eq. (\ref{Eres}). 
A cut on the minimal
reconstructed photon energy $E_{\gamma}>E_{\min}=0.3$~GeV is applied to ensure that energy distributions in the model and data are similar (see section \ref{sec:Pi0data}). 

\begin{figure}[h]
\unitlength\textwidth
\centering
\includegraphics[width=0.5\linewidth]{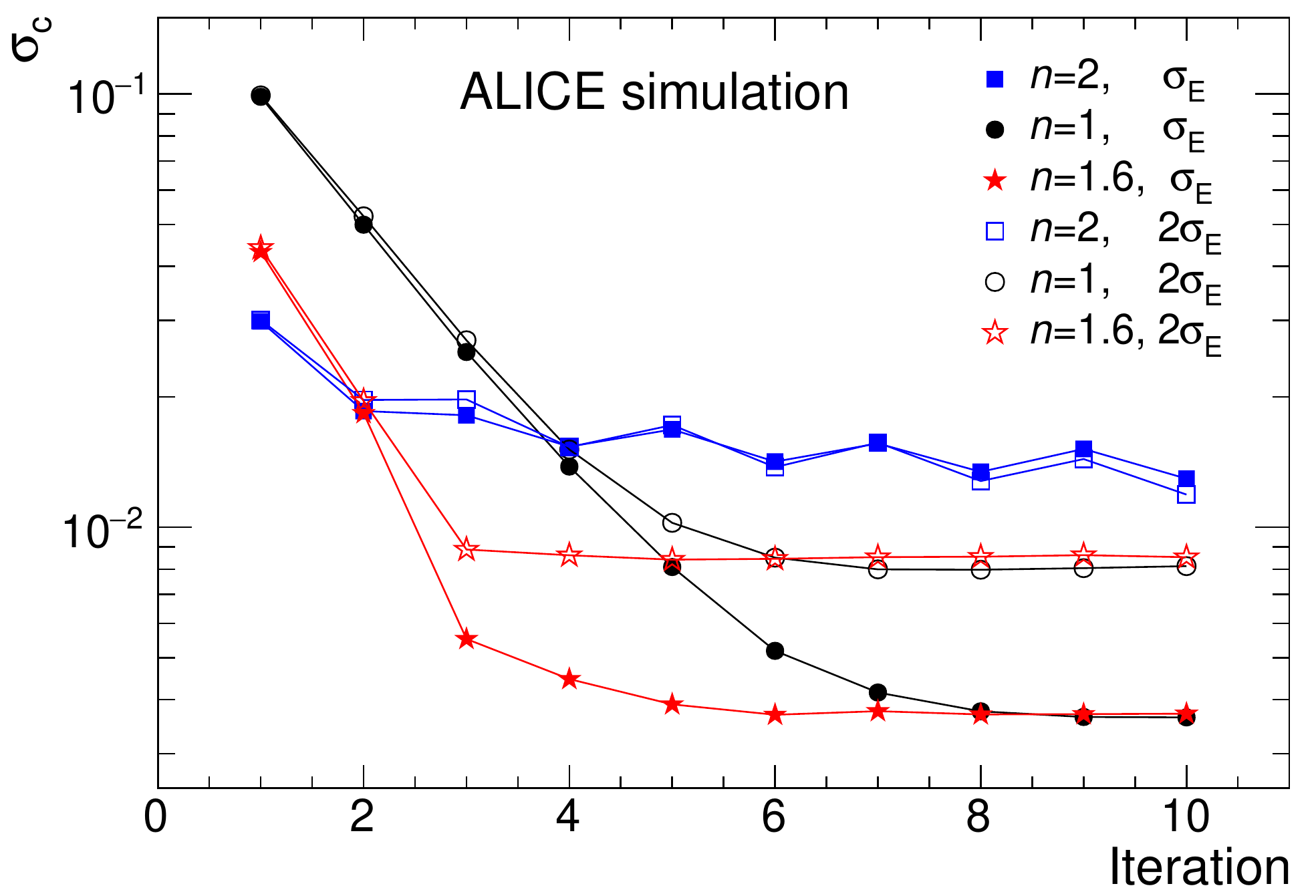}
\caption{
\label{fig:ToyIterations} [Color online]
Study using a toy Monte Carlo simulation of the convergence of the iterative calibration procedure based on equalization of the $\pi^0$ peak position. 
The residual de-calibration $\sigma_{\rm c}$ is shown as a function of the iteration number. Two values of calorimeter energy resolution are considered, standard ($\sigma_{E}$) and twice as poor ($2\sigma_{E}$). }
\end{figure}

Figure~\ref{fig:ToyIterations} shows the dependence of a residual de-calibration $\sigma_{\rm c}$, defined as the RMS of the difference 
between estimated and true calibration coefficients
$\alpha_{i}-\alpha_{i}^{\rm true}$ for all cells of the toy simulation, versus iteration number.
All calibration procedures start from the same initial de-calibration of cells and use the same sample of $\pi^0$ mesons. 
The final precision of the calibration depends on the accuracy of the reconstructed pion peak position for a typical cell, which in turn depends on the peak width 
(defined by the energy and position resolution) and the available statistics. 
In the model, the number of the simulated pions is defined by a
requirement to have $10^3$ reconstructed photons per cell after a
$\pT$ cut of $1.7~\GeVc$ on the reconstructed photon pairs. This corresponds to the calibration using real data, as described in section \ref{sec:Pi0data}.

To study the dependence of the final calibration
accuracy on the energy resolution, the default energy resolution of
the toy calorimeter was decreased by a factor of 2; these simulations are marked as $2\sigma_{\rm E}$. For powers $n<2$, the residual de-calibration stabilizes at values corresponding to the final precision of the calibration. 
In the case of $n=2$, the residual de-calibration rapidly decreases at the first iteration, but after 2$-$3 iterations start to oscillate with much slower convergence compared to values of $n<2$.

\begin{figure}[h]
  \unitlength\textwidth
  \includegraphics[width=0.49\linewidth]{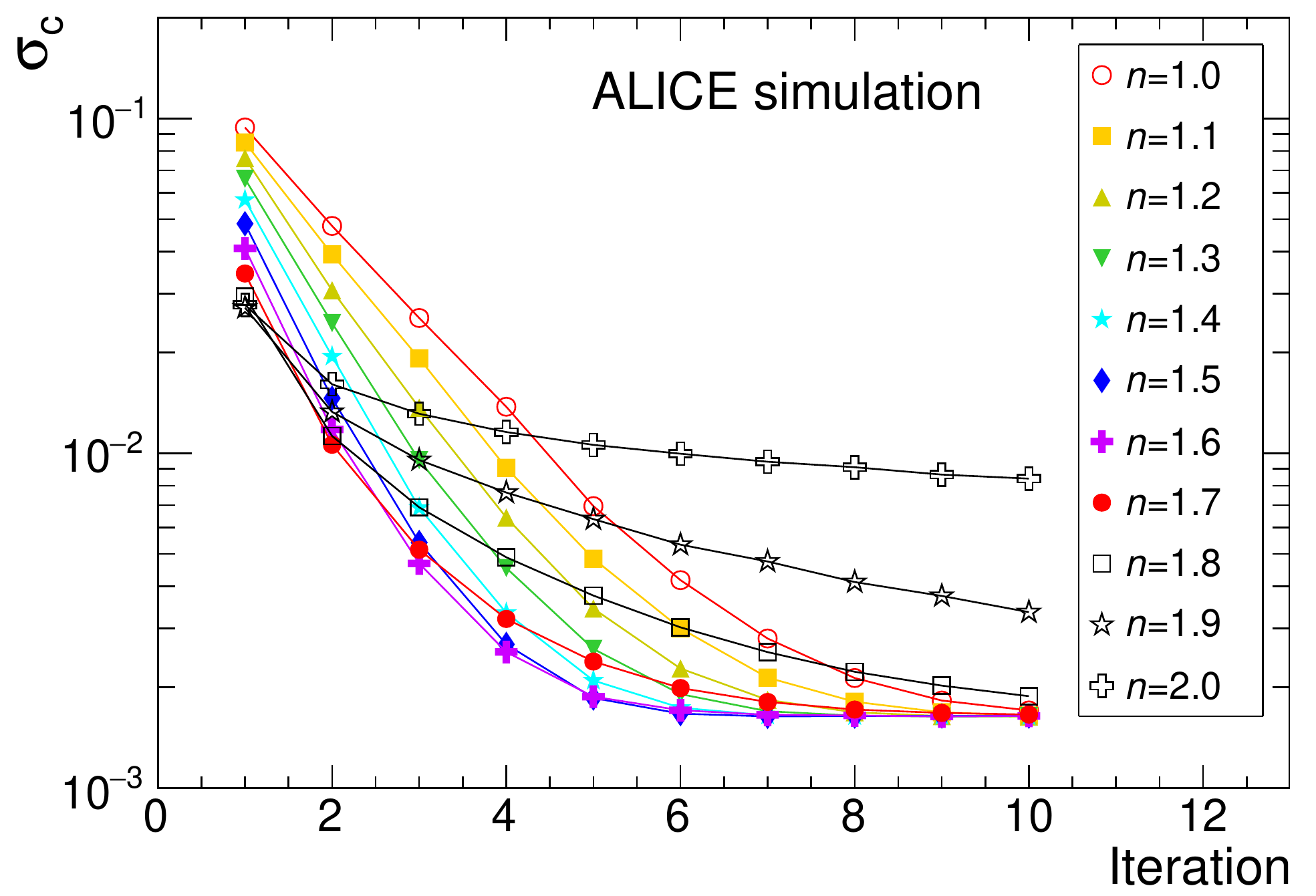}
  \hfil
  \includegraphics[width=0.49\linewidth]{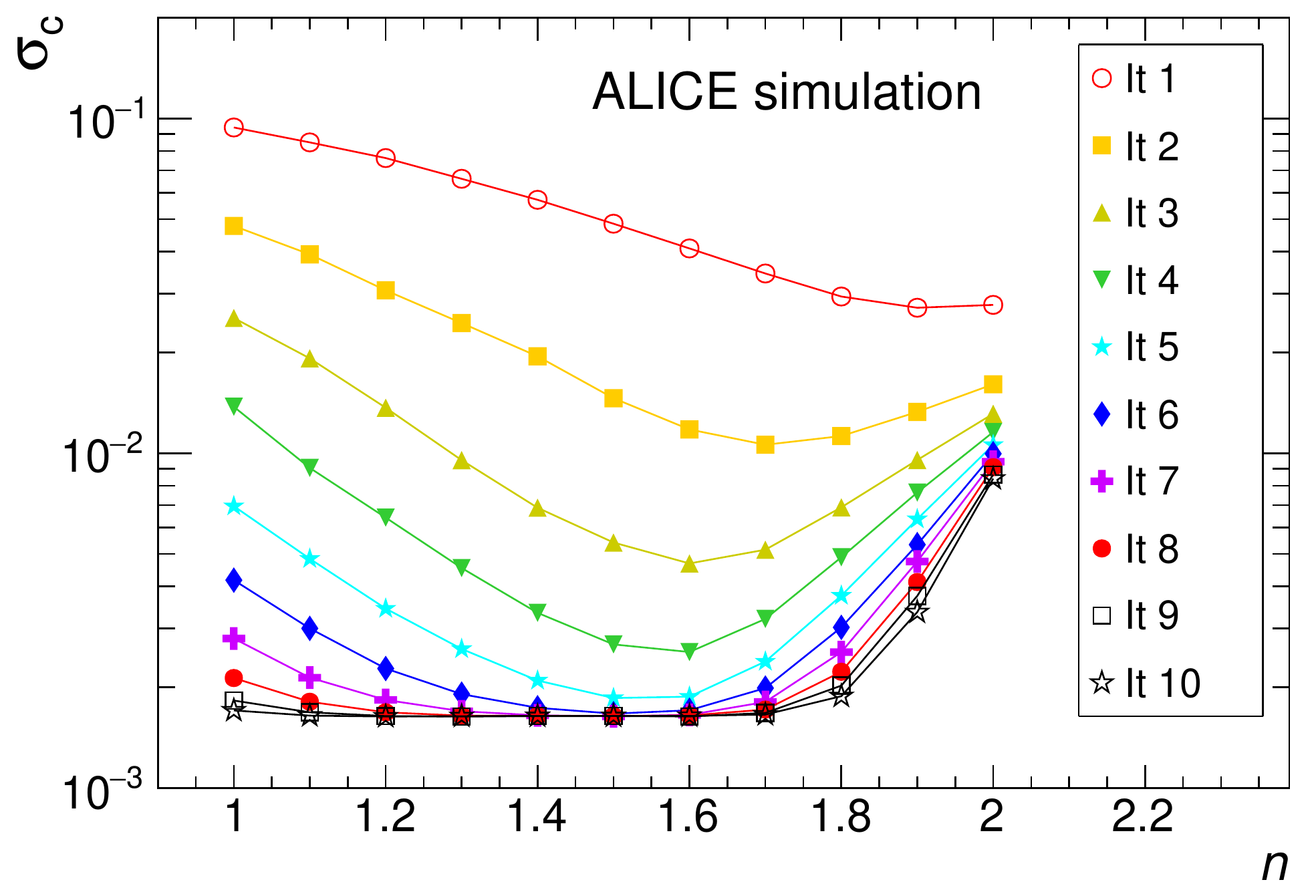}
  \caption{
    \label{fig:IterationsN} [Color online]
    Left: Residual de-calibration in the toy model simulation
    with default energy resolution versus iteration number for several
    values of power $n$. Right: Residual de-calibration versus
    power $n$ for several iterations.}
\end{figure}

In order to find the optimal value of $n$, the RMS of the de-calibration distribution is studied as a function of iteration number for several values of $n$, see 
Fig.\ \ref{fig:IterationsN} (left), and versus $n$ for several iterations (right). 
For large values of $n$, only a few iterations are needed to reach saturation. 
However, better accuracy is obtained for lower values of $n$.
Since each iteration, in an analysis with real data, is very time-consuming we chose a value of $n=1.6$ in the next analysis steps, which provides the best accuracy after 2$-$3 iterations.

\subsubsection{$\pi^0$ calibration using pp collision data}
\label{sec:Pi0data}

The procedure described above is used in the final step of the
calibration of the PHOS detector. The calibration is performed using physics data from pp collisions at
$\sqrt{s}=13$ TeV recorded in 2017. The sample contains $7.7\cdot 10^8$ minimum bias (MB)
events and $5\cdot 10^7$ events recorded with the PHOS L0 trigger~\cite{Wang:2011zzd,Kral:2012ae}, corresponding to an integrated luminosity of $\mathcal{L}_{\rm int}
=12~\mbox{nb}^{-1}$  and $5.9~\mbox{pb}^{-1}$, respectively. 
The calibration correction is only applied to the central cell of
a cluster. Clusters that are close to a dead cell are not removed. Instead, the standard approach is extended to such clusters. As a result, the shower leakage to bad cells is
compensated by higher calibration coefficients in adjacent good cells. 
A set of cuts are applied: on the minimum number of cells in a cluster, $N_{\rm cells}>2$, the minimum cluster energy $E_{\rm clu}>E_{\min}=0.3$~GeV, and the dispersion cut~\cite{Alessandro:2006yt} 
\begin{equation}
D=\sum w_i((x_i-\overline{x})^2+(z_i-\overline{z})^2)/w >0.2 \,\, \mathrm{cm}^{2},
\label{disp} 
\end{equation}
where $x_i$, $z_i$ are the coordinates of the cell $i$, $\overline{x}$,
$\overline{z}$ are coordinates of the cluster center of gravity in the PHOS
plane and the weights $w=\sum w_i$, with $w_i=\max(0,\log(E_i/E_{\rm
  clu})+4.5)$ are calculated using the energy deposition in a cell $E_i$
and the total cluster energy $E_{\rm clu}$. These cuts are used to select photon clusters and reject rare events induced by hadron interactions directly in the APD which result in disproportionally high signals~\cite{Bialas:2013wra}. 
A minimum pion transverse momentum cut $\pT>1.7~\GeVc$ is imposed to reduce the combinatorial background. 

\begin{figure}[t!]
  \centering
  \includegraphics[width=0.50\hsize]{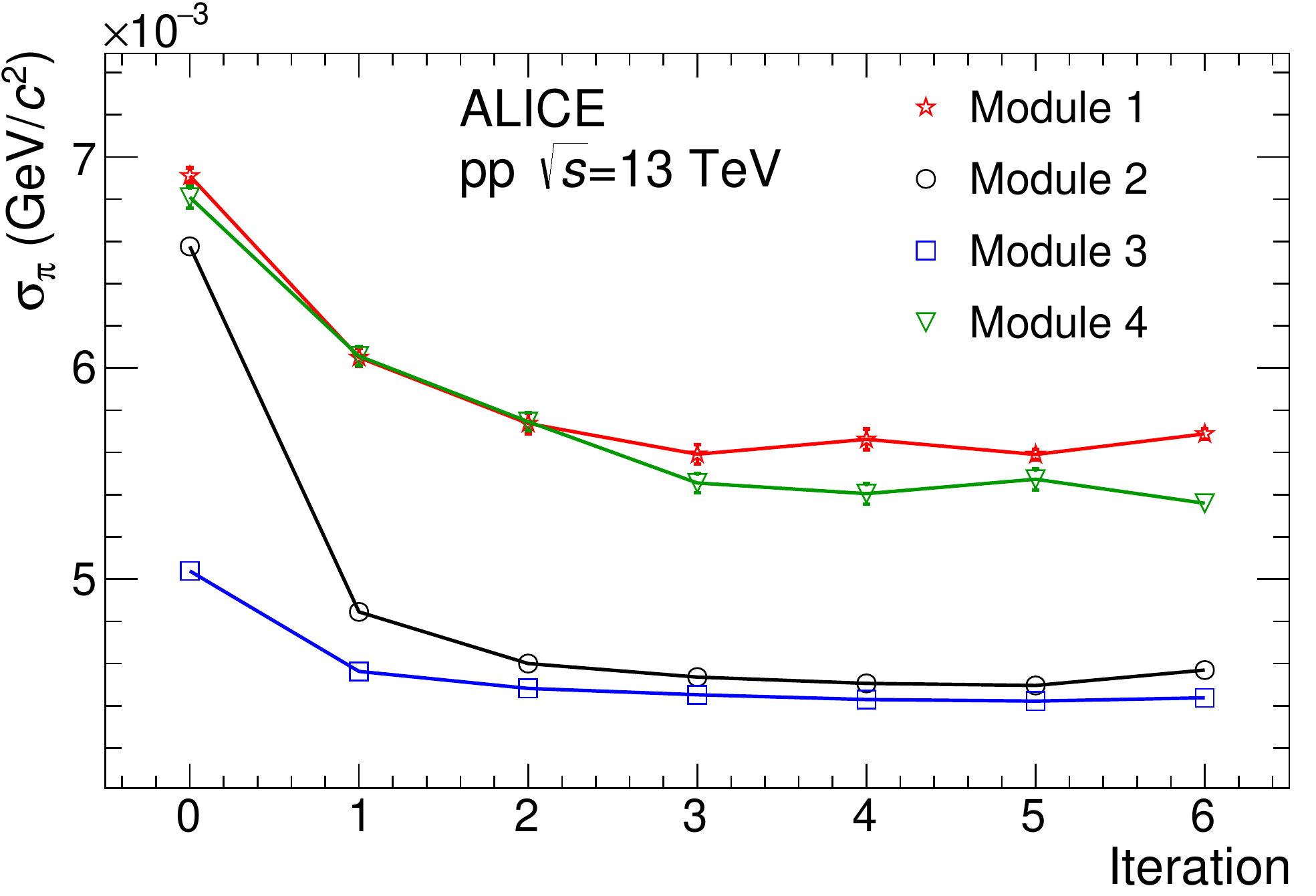}
  \caption{
    \label{fig:Iterations} [Color online]
    Dependence of the $\pi^0$ peak width on the iteration number for photon pairs with $\pT>1.7~\GeVc$ in four PHOS modules.}
\end{figure}

At each iteration the correction for the calibration coefficients is calculated using power $n=1.6$. Figure\ \ref{fig:Iterations} shows that about 3 iterations are sufficient to reach an almost final calibration. 
This is in good agreement with the predictions of the toy Monte Carlo.
The width of the peak in modules 2 and 3 is close to what is expected from Monte Carlo simulations by taking into account the PbWO$_4$ response and ideal calibration. 
In modules 1 and 4, the width is larger because of a batch of
front-end electronics cards with somewhat higher noise characteristics.

\section{Check of the energy scale}

Fixing the $\pi^0$ peak position to the PDG value is not sufficient to fix the absolute energy scale of the calorimeter. As shown in Eq. (\ref{eq:mgg}), the measured mass depends both on the cluster energy and on the opening angle. 

The reconstructed opening angle is dependent on the distance of the detector to the IP. An evaluation and check of the detector geometry is discussed in section \ref{sec:Align}. To study possible biases to the absolute energy scale, an independent cross-check was performed using the $E/p$ ratio using identified electrons.
The electrons were identified using the ALICE central tracking system, consisting of the Inner Tracking System (ITS) and the Time Projection Chamber (TPC) \cite{Alme:2010ke, Dellacasa:1999kf}, and by matching tracks with PHOS clusters.

\subsection{Calibration using identified electrons}
\label{sec:Electron}

Using electrons for the absolute energy calibration of an electromagnetic calorimeters is a widely used approach \cite{Chatrchyan:2013dga}.
In the PHOS, electrons and photons effectively deposit all their energy in the
calorimeter. It is possible to compare the energy measured in the calorimeter 
with the momentum of an electron reconstructed in the tracking system
upstream of the calorimeter.
There are two advantages of this approach compared to the calibration using the $\pi^0$ mass peak. First, only single clusters are considered and no iterative procedure is necessary. Second, the method does not depend on the exact position of the calorimeter. The geometrical mis-alignment,
appearing in the calculation of the opening angle $\theta_{12}$ in the Eq. (\ref{eq:mgg}),
is not mixed with the energy calibration. The disadvantages of this method concern the limited number of reconstructed electrons and the effects of the material budget in front of the calorimeter. 
Furthermore, this method can be used as a cross-check for the calibration using the $\pi^0$ mass peak.

\begin{figure}[ht]
  \vspace*{-4pt}
  \unitlength\textwidth
  \centering
  \includegraphics[width=0.48\linewidth]{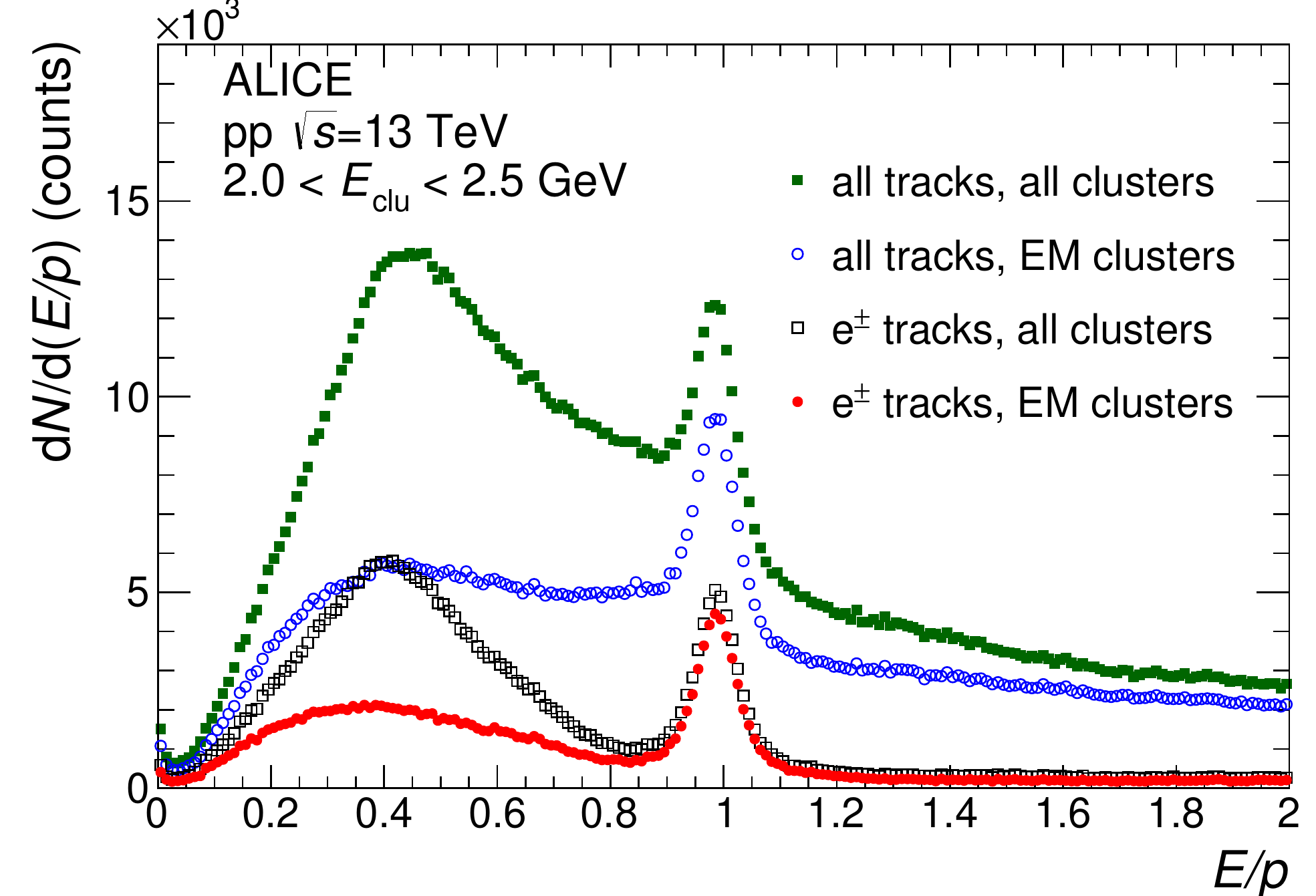}
  \hfill
  \includegraphics[width=0.48\linewidth]{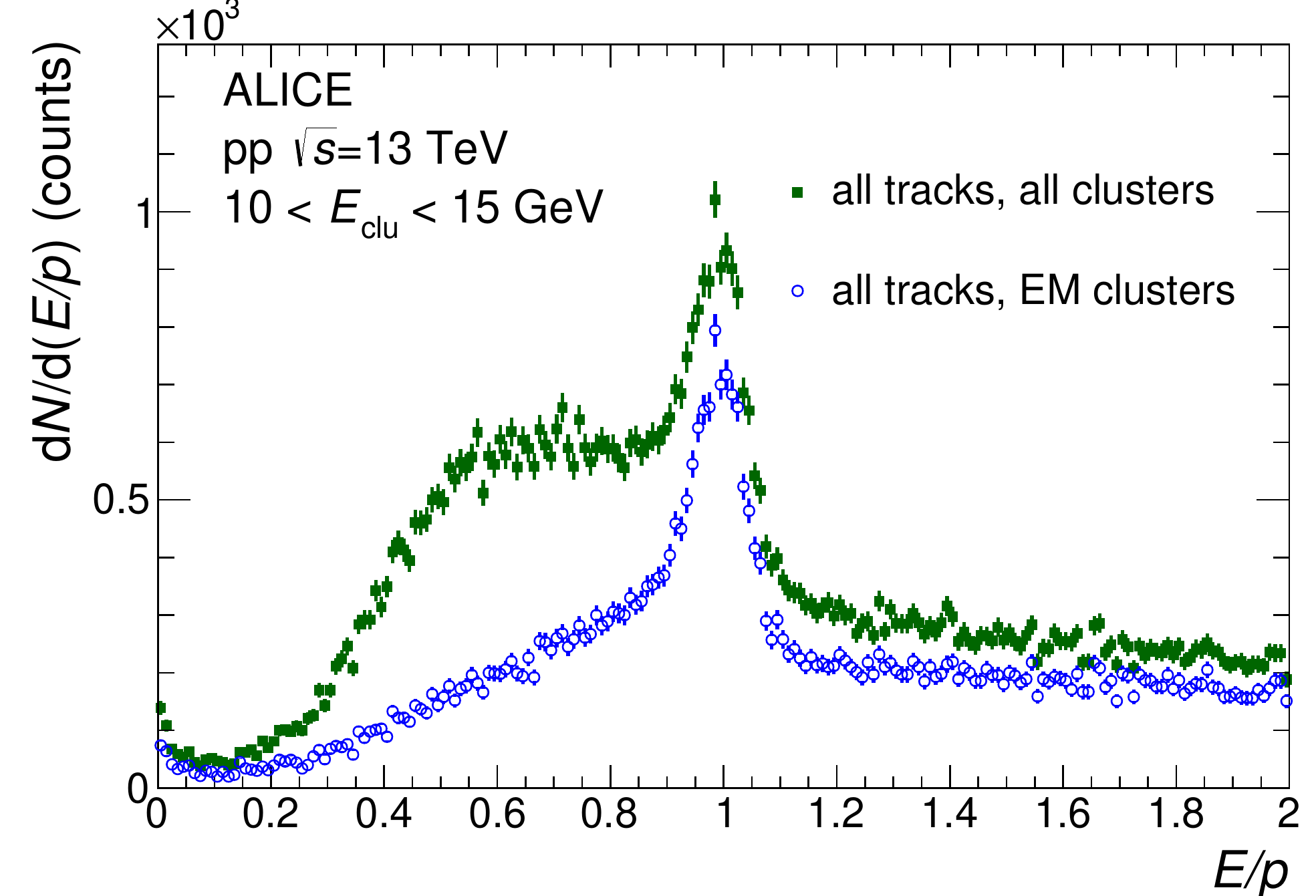}
  \vspace*{-7pt}
  \caption{
    \label{fig:ElectronEp} [Color online]
    Distribution of the cluster energy to track momentum, $E/p$ ratio, for
    two ranges of cluster energies $E_{\rm clu}$ in one PHOS module. 
    A peak around unity due to the electron contribution is visible.}
\end{figure}

The $E/p$ calibration was carried out using pp collisions at $\sqrt{s}=13$ TeV
in 2017, i.e. the same data set as that used for the $\pi^0$ calibration. Charged tracks were reconstructed with the ALICE central tracking system. Figure \ref{fig:ElectronEp} shows the $E/p$ ratio distribution for two ranges of cluster energies in a PHOS module. $E$ is the energy of the cluster in the calorimeter and $p$ is the reconstructed track momentum.
Electrons can be identified in the region around $E$/$p$ = 1 independently from the d$E$/d$x$ method provided by the tracking system. 

An optional cut is applied on the cluster dispersion, using Eq. (\ref{disp}), that corresponds to the expected electromagnetic
shower transverse size. These $E$/$p$ distributions are marked as `EM clusters' in Fig.\ \ref{fig:ElectronEp}. This reduces the background from hadrons both at low and high $\pT$, and keeps the efficiency close to 100\%.

To improve the accuracy of the peak reconstruction, the signal-to-background ratio was further improved by selecting electrons that were
identified through their specific ionization energy loss, d$E$/d$x$, in
the TPC \cite{Alme:2010ke,Adam:2015qda}. This method is efficient at low $\pT$.  
However, in the region of relativistic rise for pions, $\pT\gtrsim 1$~\GeVc, a separation of pions and electrons becomes increasingly difficult. The available statistics is not sufficient 
to perform a channel-by-channel calibration for all 12\,544 channels with good accuracy. 
%
\begin{figure}[h!]
\unitlength\textwidth
\centering
\includegraphics[width=0.49\linewidth]{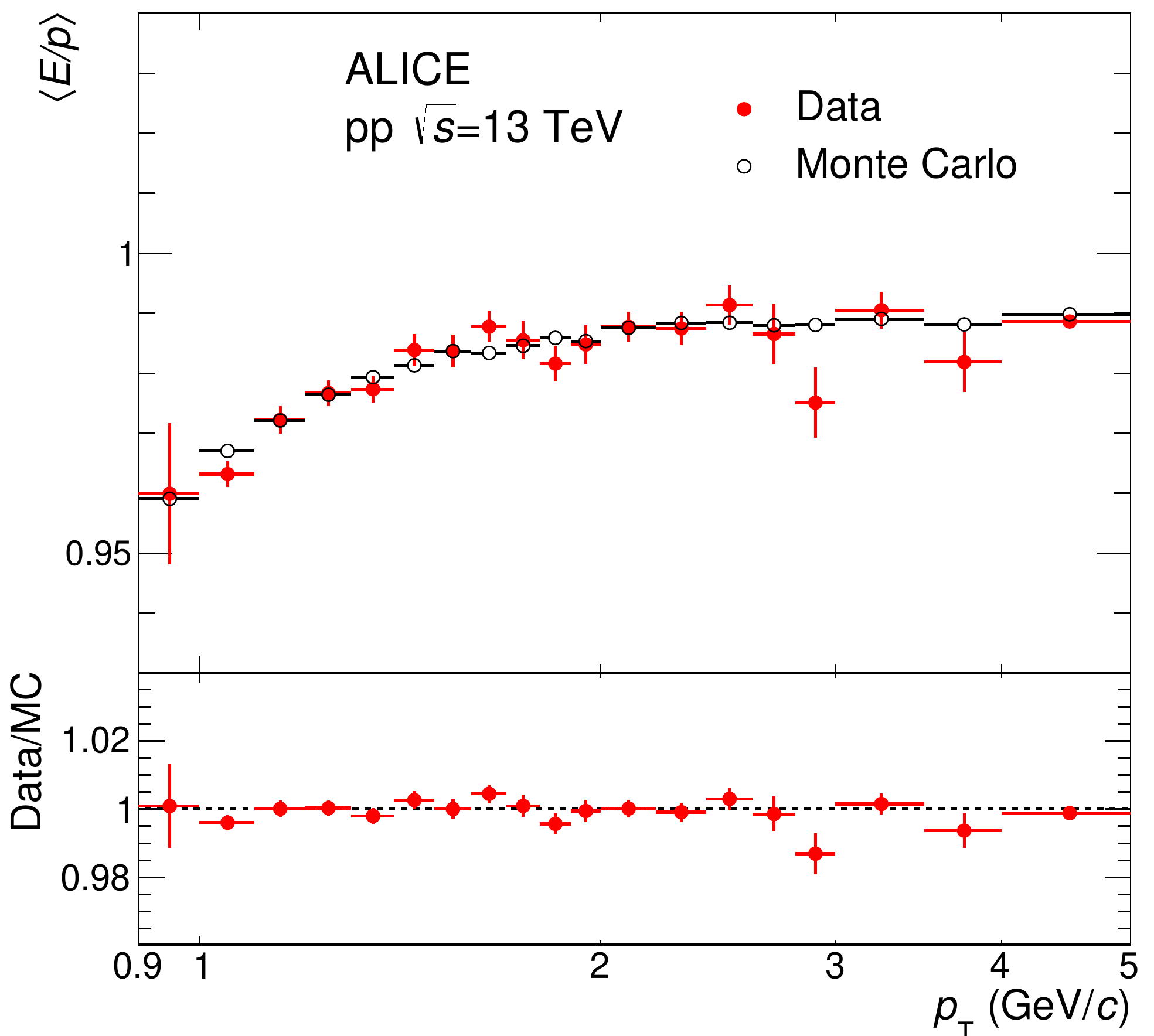}
\hfill
\includegraphics[width=0.49\linewidth]{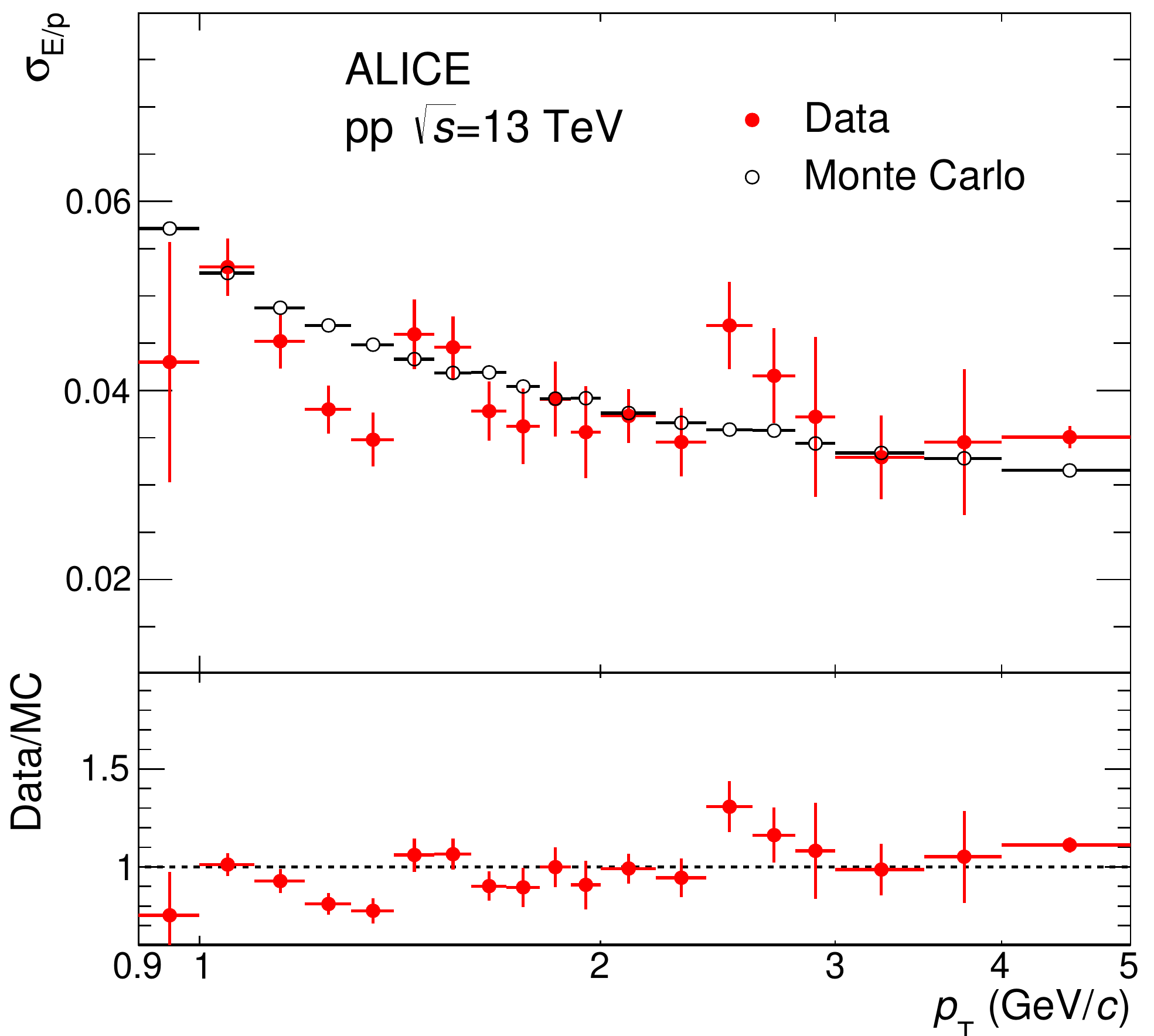}
\caption{
\label{fig:ElectronEpMean} [Color online]
Mean (left) and width (right) of the $E/p$ peak position in data and MC for electron candidates.}
\end{figure}
%
Figure \ref{fig:ElectronEpMean} shows the $E/p$ peak position and the peak width, after fitting the
$E/p$ distributions with the dispersion cut applied, as a function of cluster energy. The data are from the two middle PHOS modules. These modules have the best energy resolution. 
Note that the non-linearity corrections, discussed in section \ref{sec:Nonlin} are applied in this analysis for comparisons with Monte Carlo simulations.
At high $\pT$, the mean is close to unity, but gradually decreases towards smaller $\pT$, reflecting an increased relative energy loss of lower energy electrons. 
Figure \ref{fig:ElectronEpMean} also shows the results from Monte Carlo simulations with the PYTHIA8 event generator \cite{Sjostrand:2014zea}
using the standard ALICE software framework for the analysis of real data. The simulation includes a remaining small mis-calibration describing an inaccuracy of our calibration to reproduce the $\pi^0$ mass peak
position and width and their dependence on $\pT$. 
The agreement is better than $\sim 0.2$\% providing an independent estimate of the absolute energy scale precision in the PHOS.

\subsection{Geometrical alignment}
\label{sec:Align}

The precise measurement of the distance between the IP and the calorimeter surface, $R$, is a
difficult task because of the detectors installed in front of PHOS.
Uncertainties in the measurement of $R$ directly translate to uncertainties in
the energy scale. 

Equation (\ref{eq:mgg2}) shows the dependence on $R$ and the distance between the clusters, $L_{12}$, in the calorimeter for the calculation of the two-photon invariant mass:
\begin{equation}
\label{eq:mgg2}
 m_{\gamma\gamma}=2\sqrt{E_1E_2}|\sin(\theta_{12}/2)|\approx \sqrt{E_1E_2} \frac{L_{12}}{R},
\end{equation}

The alignment of the PHOS was measured via the
photogrammetry procedure \cite{Photogrammetry}. In addition, an independent estimate of the PHOS alignment is performed by 
matching tracks reconstructed in the tracking system with 
clusters in PHOS. To study the alignment it is convenient to use the local coordinate system of the PHOS module where $z$ is the coordinate along the beam and $x$ is the coordinate perpendicular to the beam direction. The alignment in the $z$ and $x$ directions is straightforward, unlike checks for the radial distance.

\begin{figure}[ht]
  \centerline{
    \includegraphics[width=0.6\textwidth]{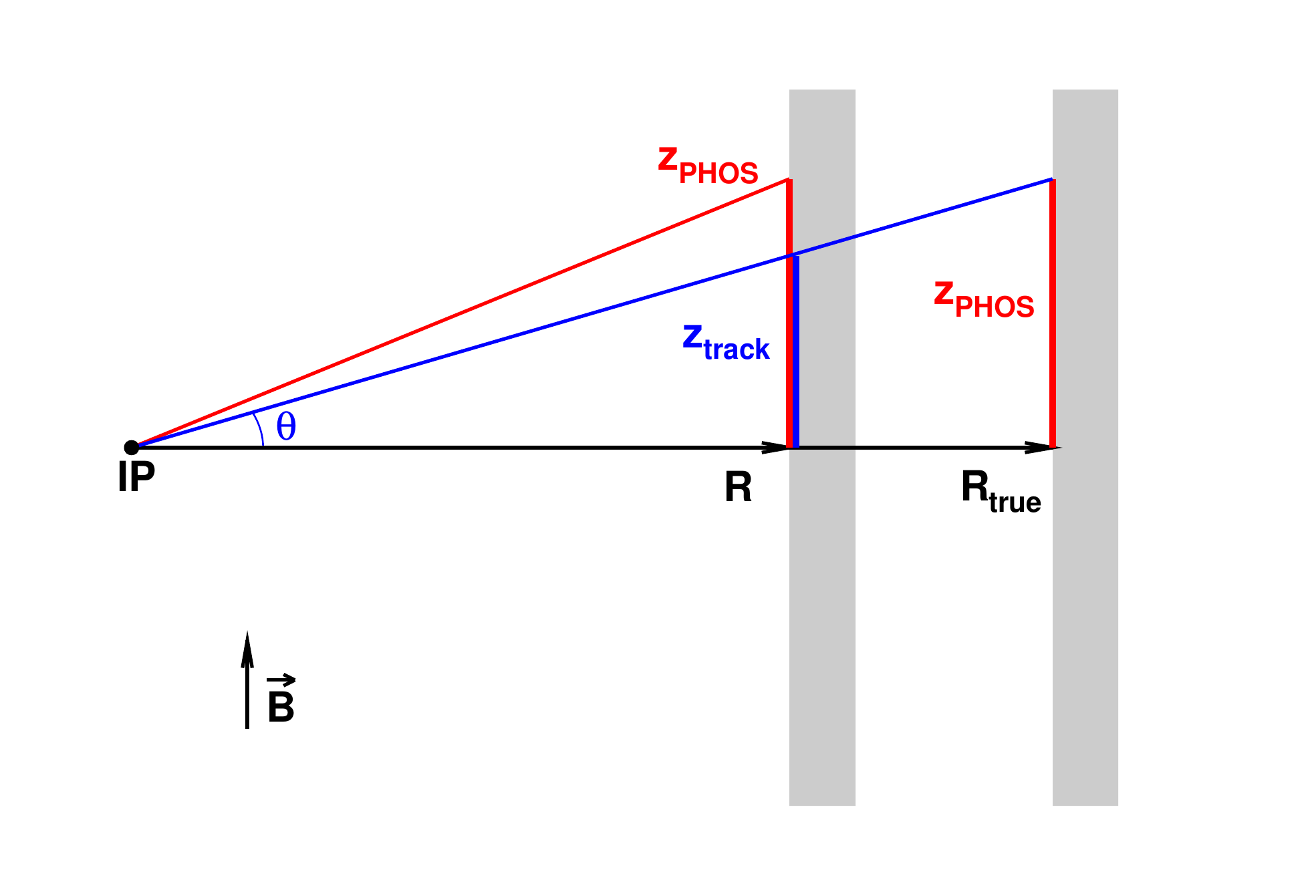}
  }
  \caption{[Color online] An illustration of the dependence of $\langle dz \rangle$, from equation (\ref{eq:dz-def}), with $z$, in a
radially shifted detector. The magnetic field of 0.5 T is along the $z$ direction.}
  \label{fig-Misal-scheme} 
\end{figure}

Figure \ref{fig-Misal-scheme} shows the geometry and variables used to establish the radial distance of the PHOS from the IP. The difference between the $z$ coordinate of the reconstructed cluster position in the calorimeter, $z_{\rm PHOS}$, and the point of the track extrapolated to the surface of the calorimeter, $z_{\rm track}$, through the ratio of true ($R_{\rm true}$) and expected ($R$) radial distances is:
\begin{equation}\label{eq:dz-def}
    dz = z_{\rm PHOS}-z_{\rm track} = z_{\rm PHOS} -  R\,\tan \theta = z_{\rm PHOS}\left( 1- \frac{R}{R_{\rm true}}\right).
\end{equation}
%

In this analysis, the depth of the shower maximum for a photon is used as
a reference point \cite{Alessandro:2006yt}. A correction for this depth is introduced to the cluster center of gravity so that the $x$ and $z$ coordinates correspond to those of the photon at the front surface of PHOS. 
In contrast to photons and electrons, because of the large nuclear interaction length of the EM calorimeter, 
the center of gravity of a hadronic shower 
is almost uniformly distributed in the depth of the calorimeter and therefore hadronic tracks are not suitable for such calibration.
Electron showers reach their shower maximum about one unit in radiation length
$X_{0}$ earlier than photons. 
The difference between the photon and electron cluster coordinate in the $z$
direction can be written as:
\begin{equation}\label{e}
  \delta z_{\rm e} = -X_{0}\sin\left(\arctan \frac{z_{\rm PHOS}}{R_{\rm true}}\right).
\end{equation}

\begin{figure}[htb]
  \includegraphics[width=0.48\textwidth]{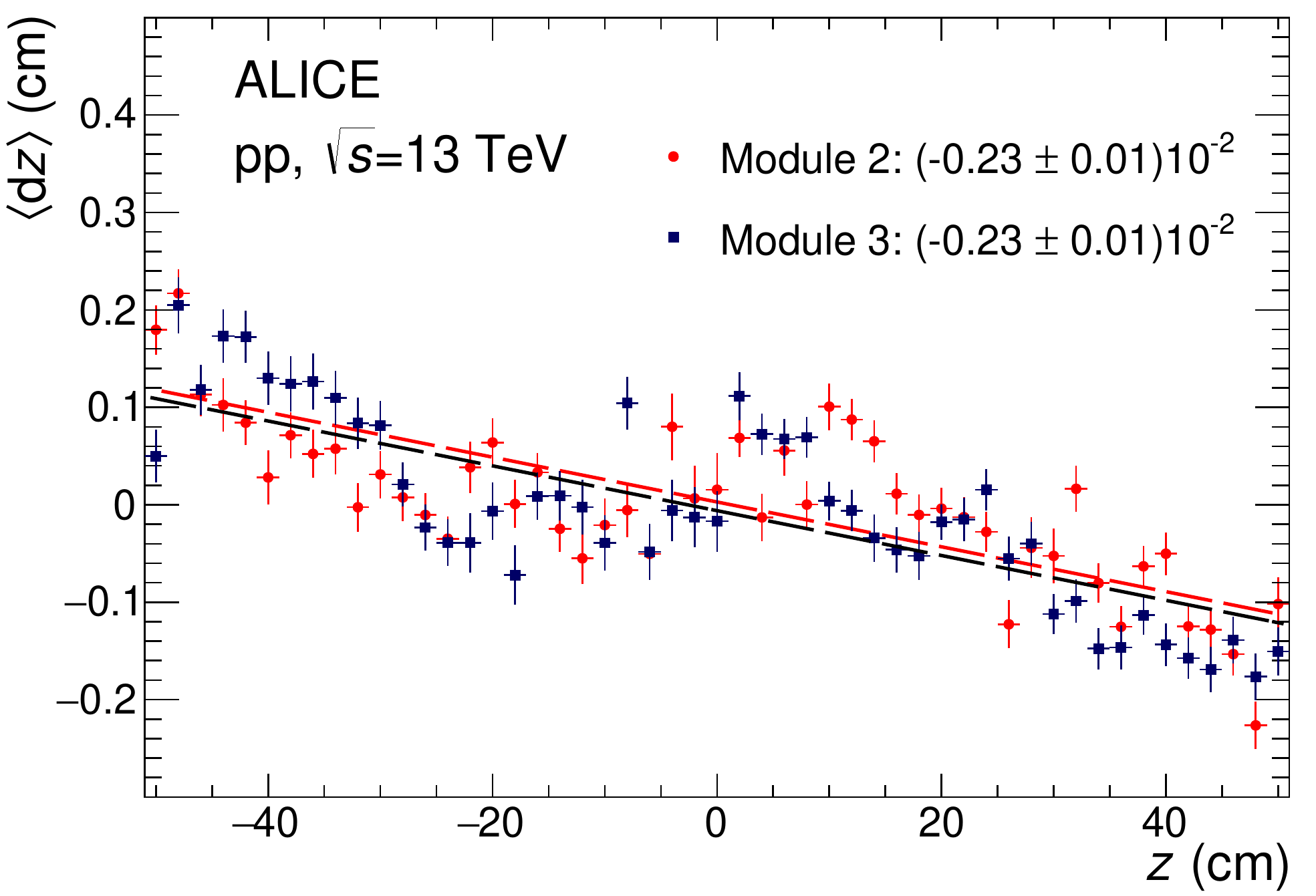}
  \hfil
  \includegraphics[width=0.48\textwidth]{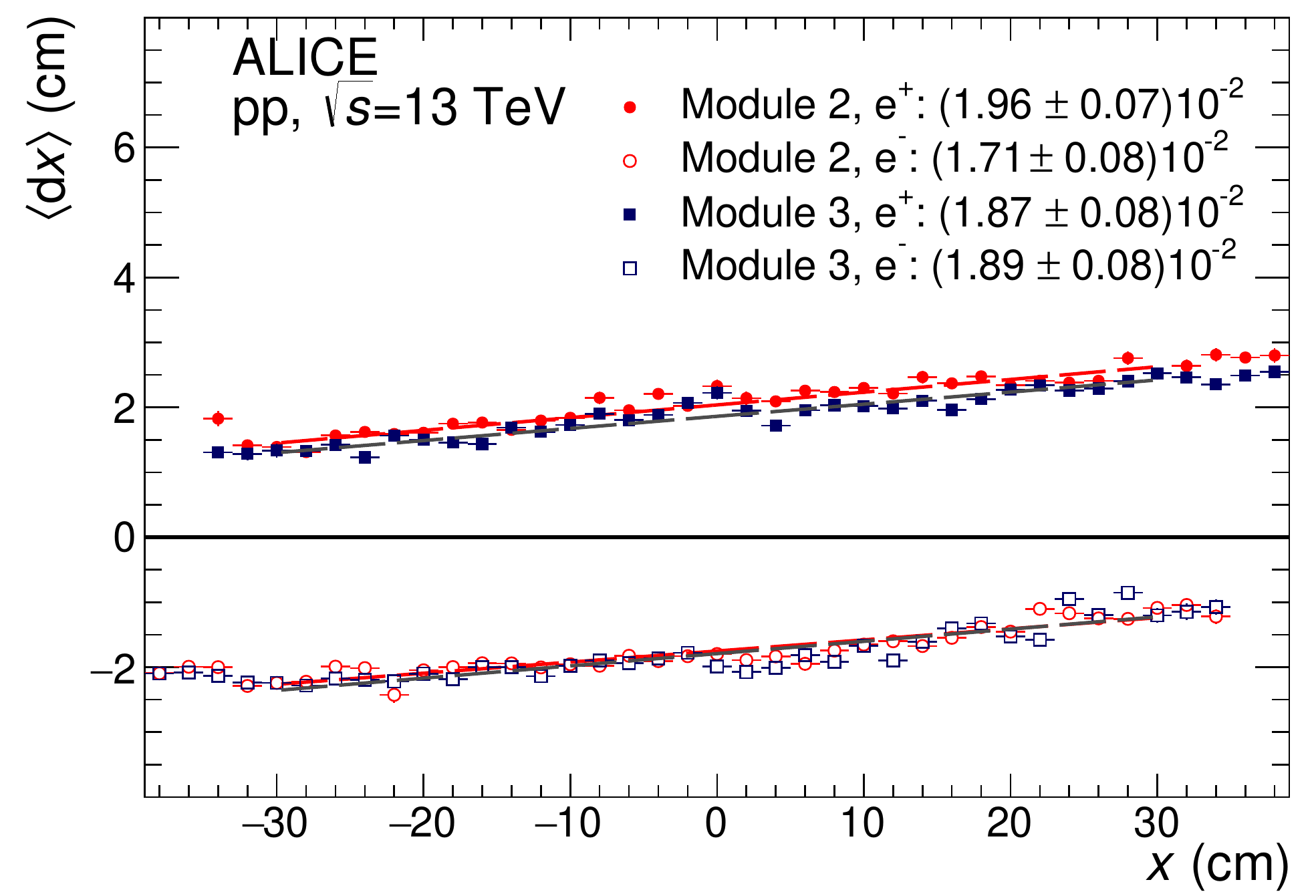}
  \caption{ [Color online] Dependence of the mean distance between track extrapolation to the PHOS surface and cluster position in the cluster coordinate on the PHOS plane along (left) and perpendicular
(right), to the beam and magnetic field direction. In the left plot contributions of electrons and positrons are combined. The dependencies are fitted with linear functions and the resulting slopes are shown in both legends.}
  \label{fig-Misal}
\end{figure}

Figure \ref{fig-Misal} (left) shows the $\langle dz \rangle$ versus $z$ dependence. The data from the two modules are very similar,
with the same slope of $-0.23\cdot 10^{-2}$. There are some oscillations around the linear dependence.
The slope is slightly larger than the expected slope, $B_{\rm e}$, from Eq. (\ref{e}), of $-0.19\cdot 10^{-2}$.
This difference corresponds to $\sim 4$ mm inward radial shift of the PHOS modules. These values were used to correct the radial PHOS position in the offline
reconstruction.

The magnetic field causes charged tracks to be bent in the radial plane,
which introduces complications in the $\langle dx \rangle$ versus $x$ analysis that are not present for the $\langle dz \rangle$ versus $z$ analysis.
Figure \ref{fig-Misal} (right) shows the $\langle dx \rangle$ versus $x$ dependence. The data for positive and negative charges have similar slopes, but opposite offsets, because of the track bending in the magnetic field. This results in different incident angles, for electrons and positrons, with respect to photons. These angles strongly depend on the particle $\pT$, making
this analysis much more complicated than the $\langle dz \rangle$ versus $z$ study. Therefore, only the $\langle dz \rangle$ versus $z$ data
are used in the final PHOS alignment procedure.

\section{Estimate of the energy nonlinearity correction}

\label{sec:Nonlin}

There are several effects that may influence the linearity of PHOS
energy measurement. At low energies, light attenuation in the crystals, electronic noise, electronic thresholds and amplitude digitization are important. At high energies, shower leakage contributes to a nonlinear response.  
For the physics analysis it is sufficient to reproduce the observed nonlinearity of the detector in the Monte Carlo simulations, but practically, it is more convenient to correct real data for the nonlinearity in order to reduce the mass resolution of a neutral meson peak in wide $\pT$ bins.


The nonlinearity is corrected through a recalculation of the
cluster energy $E$ by the following parameterization:
\begin{equation}
 E_{\rm corr} = \left \{ 
 \begin{array}{ll}
 aE + b \sqrt{E} + c + d/\sqrt{E} + e/E, & E\le E_0 \\
 \alpha E + \beta \sqrt{E},    & E>E_0 \\
\end{array}
\right .
 \label{fitNL}
\end{equation}
where free parameters $a, b, c, d, e, E_0$ are chosen to provide a
\pT-independent reconstructed neutral pion mass $m_{\pi^0}$ in pp
collisions at $\sqrt{s}=13$ TeV and parameters $\alpha$ and $\beta$
are fixed to ensure a smooth function at the point $E=E_0$.

\begin{figure}[h]
  \unitlength\textwidth
  \centering
  \includegraphics[width=0.48\linewidth]{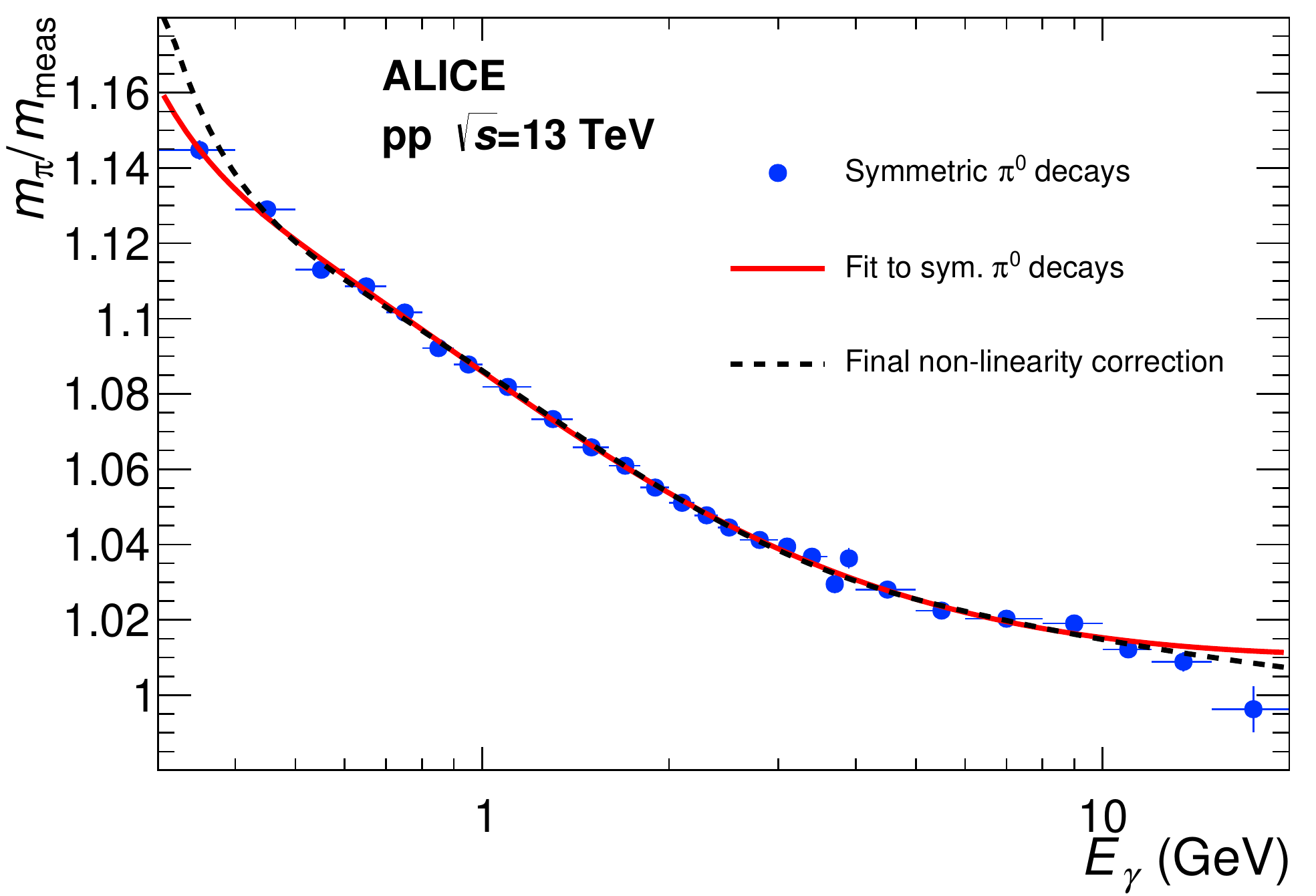}
  \caption{
    \label{fig:NonlinAlpha} [Color online]
    Estimation of PHOS nonlinearity using symmetric $\pi^0$ decays
    defined by
    $|E_{\gamma,1}-E_{\gamma,2}|<0.05(E_{\gamma,1}+E_{\gamma,2})$. Data
    fit with function (\ref{fitNL}). The final tuned nonlinearity is
    shown with a dashed curve. }
\end{figure}

Figure \ref{fig:NonlinAlpha} shows the ratio of the PDG $\pi^0$ mass to the measured $\pi^0$ peak position as a function of mean photon energy, $E_\gamma$. The data were restricted to symmetric $\pi^0$ decays with
$|E_{\gamma,1}-E_{\gamma,2}|<0.05(E_{\gamma,1}+E_{\gamma,2})$. A fit with the function $E_{\rm corr}(E)/E$ (Eq. \ref{fitNL}) is shown by the red curve.

\begin{figure}[h]
  \unitlength\textwidth
  \centering
  \includegraphics[width=0.48\linewidth]{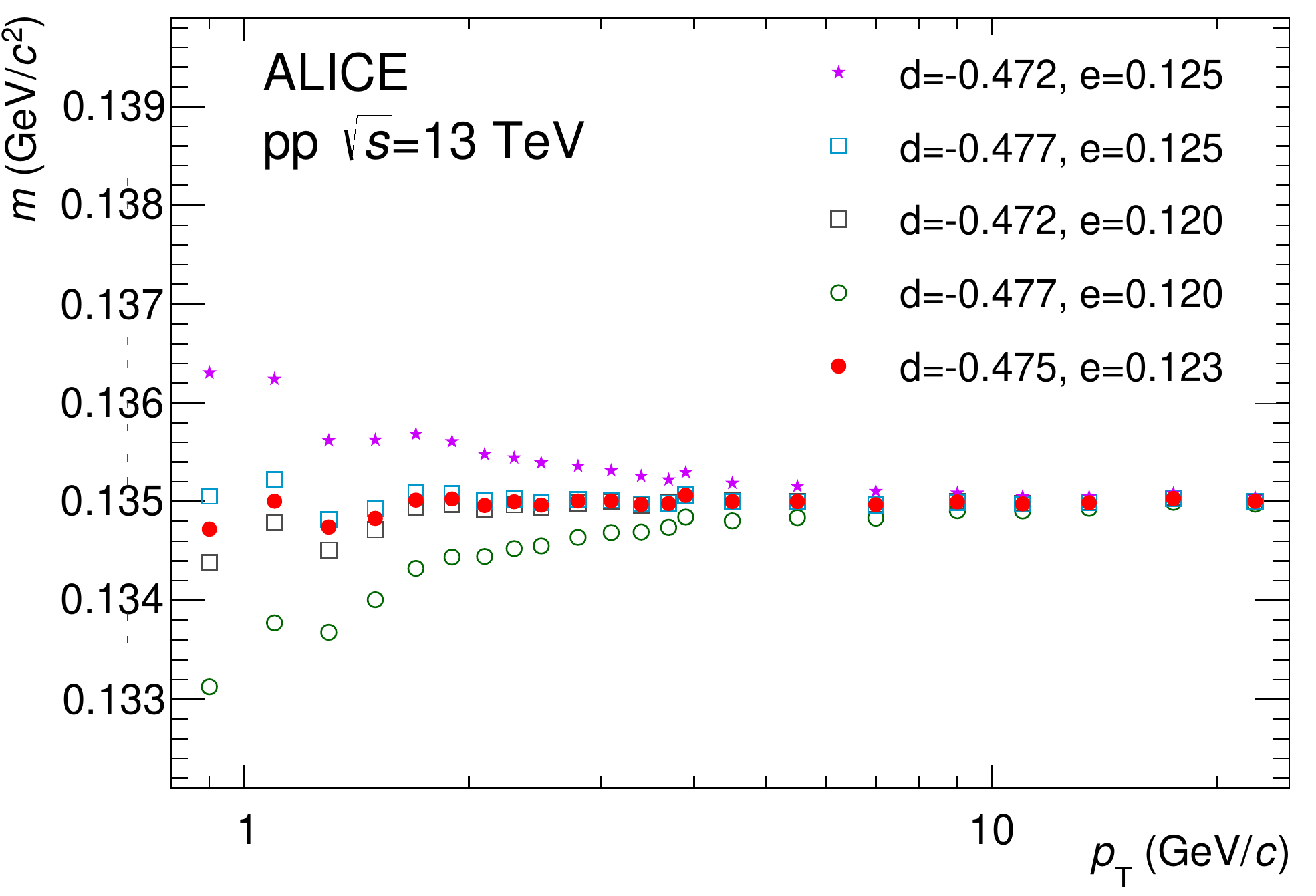}
  \hfill
  \includegraphics[width=0.48\linewidth]{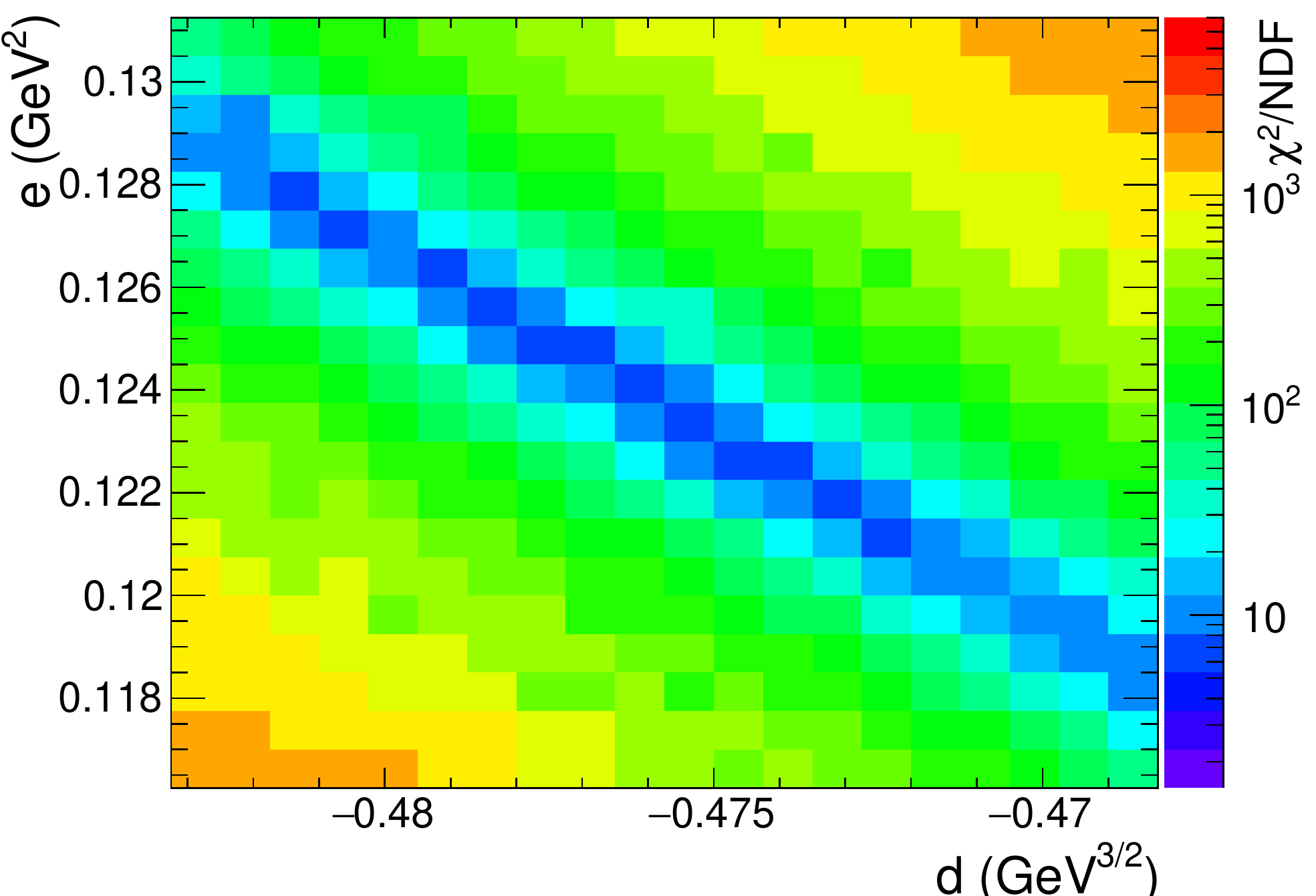}
  \caption{
    \label{fig:NonLin} [Color online]
    Left: the $\pi^0$ peak position as a function of the transverse momentum for several values of nonlinearity parameters ($d$, $e$), with default values for $a$, $b$ and $c$. Right: the deviation from a constant value of the $\pi^0$ peak position expressed in $\chi^2/NDF$ as a function of the nonlinearity parameters ($d$, $e$).}
\end{figure}

However, this method is not reliable at very low energies where systematic uncertainties for the $\pi^0$ signal extraction are large because of the limited PHOS acceptance. The same is true at high $\pT$ where photons from symmetric decays start to merge into one cluster.
To improve the nonlinearity parameterization, a set of invariant mass distributions were calculated as a function of $\pT$, without requiring symmetric decays. Each mass distribution was corrected for nonlinearity with different sets of nonlinearity parameters ($a$, $b$, $c$, $d$, $e$, $E_0$). 
Figure\ \ref{fig:NonLin} (left) shows examples of the dependence on $d$ and $e$, on the peak position versus $\pT$. 
Note that parameter $a$ sets an absolute normalization and can be factorized in this analysis. 

To find the best set of parameters, a fit of the peak $\pT$-dependence with a
constant function is performed in the range $0.6-25~\GeVc$. The resulting $\chi^2$ value for each
set of parameters is shown in Fig.\ \ref{fig:NonLin} (right). In this plot we fix optimal values of parameters $a$, $b$, $c$, $E_0$ and vary only parameters $d$, $e$. 
The optimal set, obtained by minimizing $\chi^2$, is 
($a=1.02\pm0.01$, $b=-0.2548\pm0.0005$~GeV$^{1/2}$, $c=0.648\pm0.001$~GeV, $d=-0.4743\pm0.0002$~GeV$^{3/2}$, $e=0.1215\pm0.0005$~GeV$^{2}$ and $E_0=5.17\pm0.01$~GeV).
The nonlinearity correction corresponding to this set is shown with a black dashed line in Fig.~\ref{fig:NonlinAlpha}.
This parameter set, corresponding to the filled red circles in the left plot of 
Fig.\ \ref{fig:NonLin}, is used in the offline reconstruction.

%
%

\section{Run-by-run energy calibration}
\label{sec:Run-by-run}

The light yield from the crystals, and the gain in the APDs, are strongly
temperature dependent \cite{Dellacasa:1999kd,Deiters:2000ip}. To minimize this dependency on the PHOS energy scale, the PHOS crystal matrices were thermo-stabilized to within $0.3^\circ$C. 
This temperature variation results in a change of about 0.6\% in light yield and APD gain. 
Another effect that may influence the long-term stability of the amplitude measurement in the PHOS detector is the crystal transparency dependence on the radiation dose.
A run-dependent calibration correction, common for all channels in each PHOS module, was implemented to account for all these effects. 
In order to estimate this correction, the standard calibrations and corrections were applied. For each run, the mean value of the $\pi^0$ mass peak in each module was extracted,
using only photon pairs in that module.

\begin{figure}[h]
  \unitlength\textwidth
  \centering
  \includegraphics[width=\textwidth]{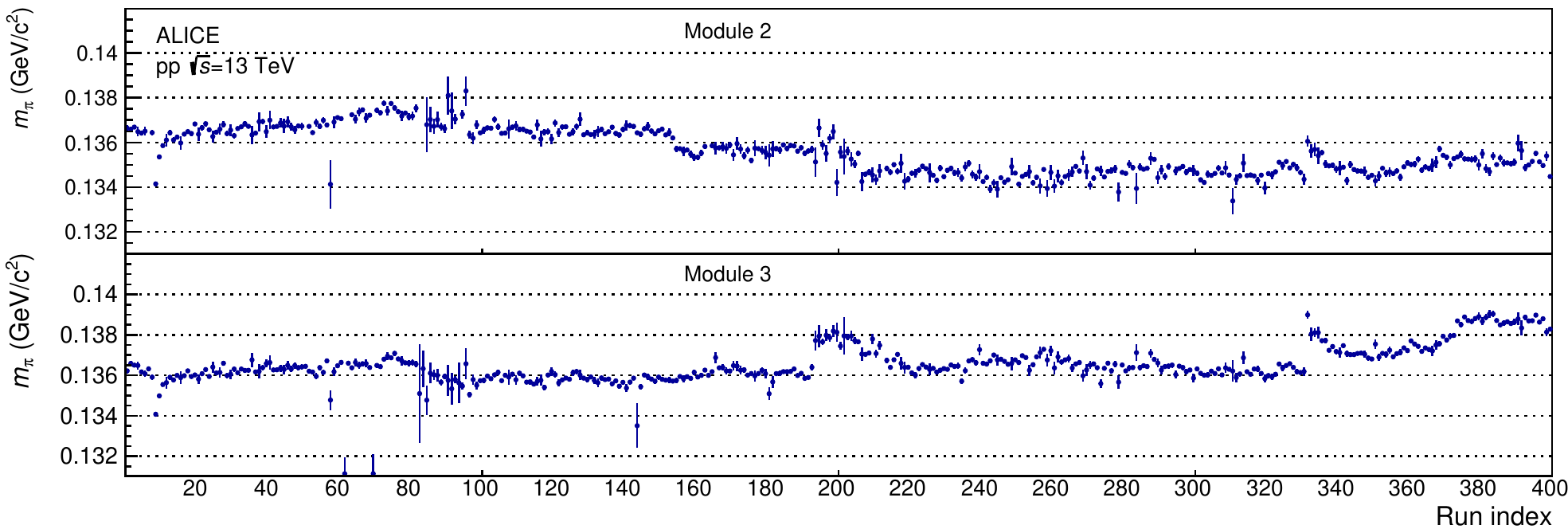}
  \caption{
    \label{fig:RunByRun} [Color online] Example of the dependence of the $\pi^0$ peak
    position on the run number for 400 sequential runs recorded during 3
    months of the 2017 data taking campaign.}
\end{figure}

The correction is calculated using the data sample collected with the PHOS L0 trigger since it has better statistics at high $\pT$, where the signal-to-background ratio is larger.
Figure~\ref{fig:RunByRun} shows the reconstructed $\pi^0$ mass peak versus run number, for 400 sequential runs, from pp collisions at $\sqrt{s}=13$~TeV, recorded during 3 months of data taking from June to September 2017, with stable running conditions. The data are for the two middle PHOS modules. These have the largest acceptance and the best energy resolution.

On average the peak position is stable to within $\sim 2$ MeV/$c^2$ in
both modules, but reveals several correlated and uncorrelated trends
in these two modules. Correlated trends are related to the powering of
the PHOS front-end electronics in both modules, and therefore to the
variation of the heat deposition and temperature of the crystal
matrix. Uncorrelated trends may have different reasons: switching on
or off isolated front-end cards, formation of ice jams in the cooling
pipes of the cooling system, etc. There is no visible global
correlated trend of a decrease of the peak position in all modules,
which would indicate a radiation damage in the crystals and a decrease of their transparency with time. The total integrated dose in the PHOS crystals accumulated during 3 years of running with pp, p$-$Pb and Pb$-$Pb beams during Run 2, is estimated to be less than 0.1 Gy. The total hadron fluence was about $2\cdot 10 ^9$~cm$^{-2}$.

In the calibration procedure the mean value of the peak position over the whole period is calculated and deviations with respect to this value are estimated. If the peak position in a module is known with uncertainty better than 1 MeV, all calibration coefficients in a module are corrected by the ratio $m_{\rm mean}/m_{\rm run}$. If a run is too short and fitting is not possible, the mean value over the whole period is used.

\section{Results of calibration}
\label{sec:Results}

\begin{figure}[h]
  \includegraphics[width=0.49\hsize]{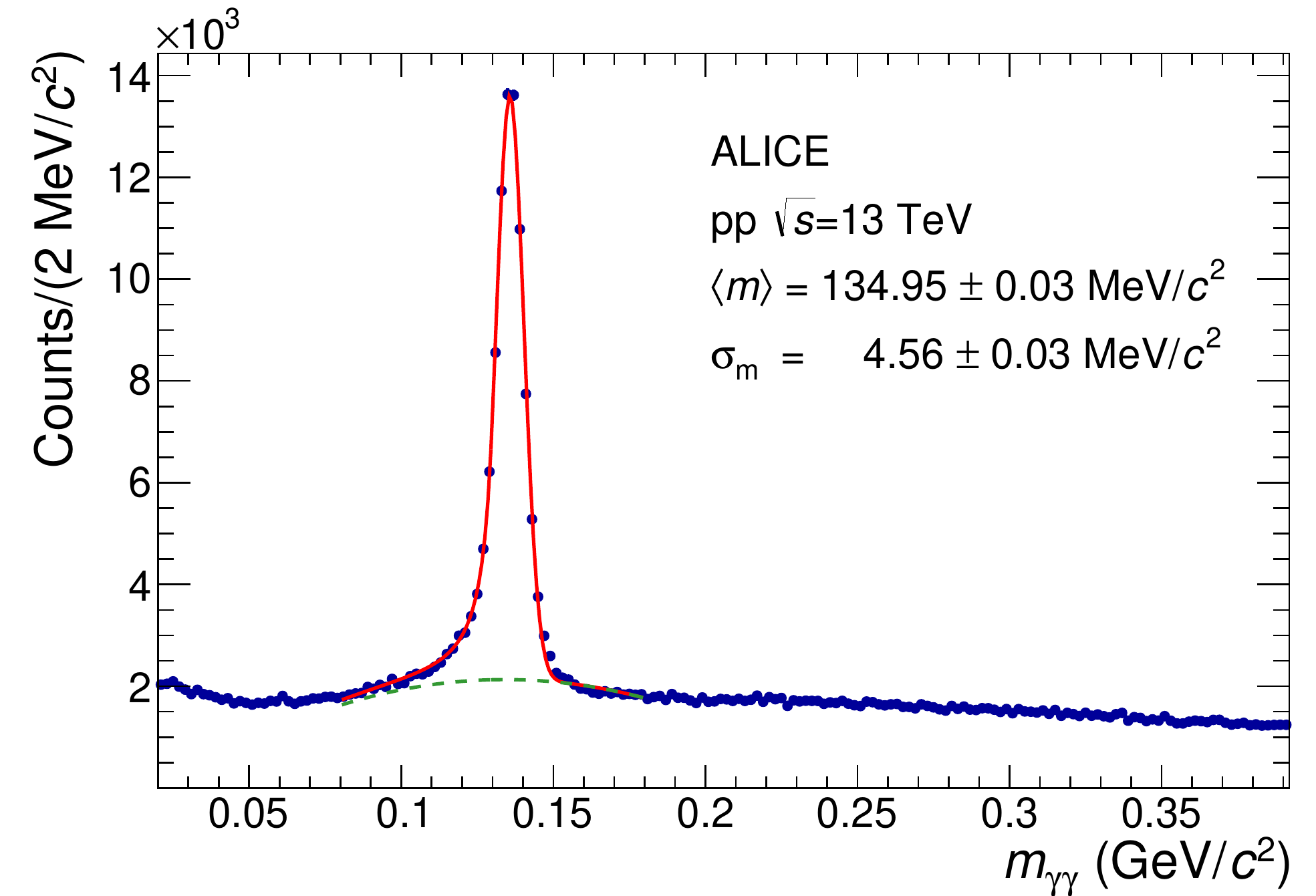}
  \hfil
  \includegraphics[width=0.49\hsize]{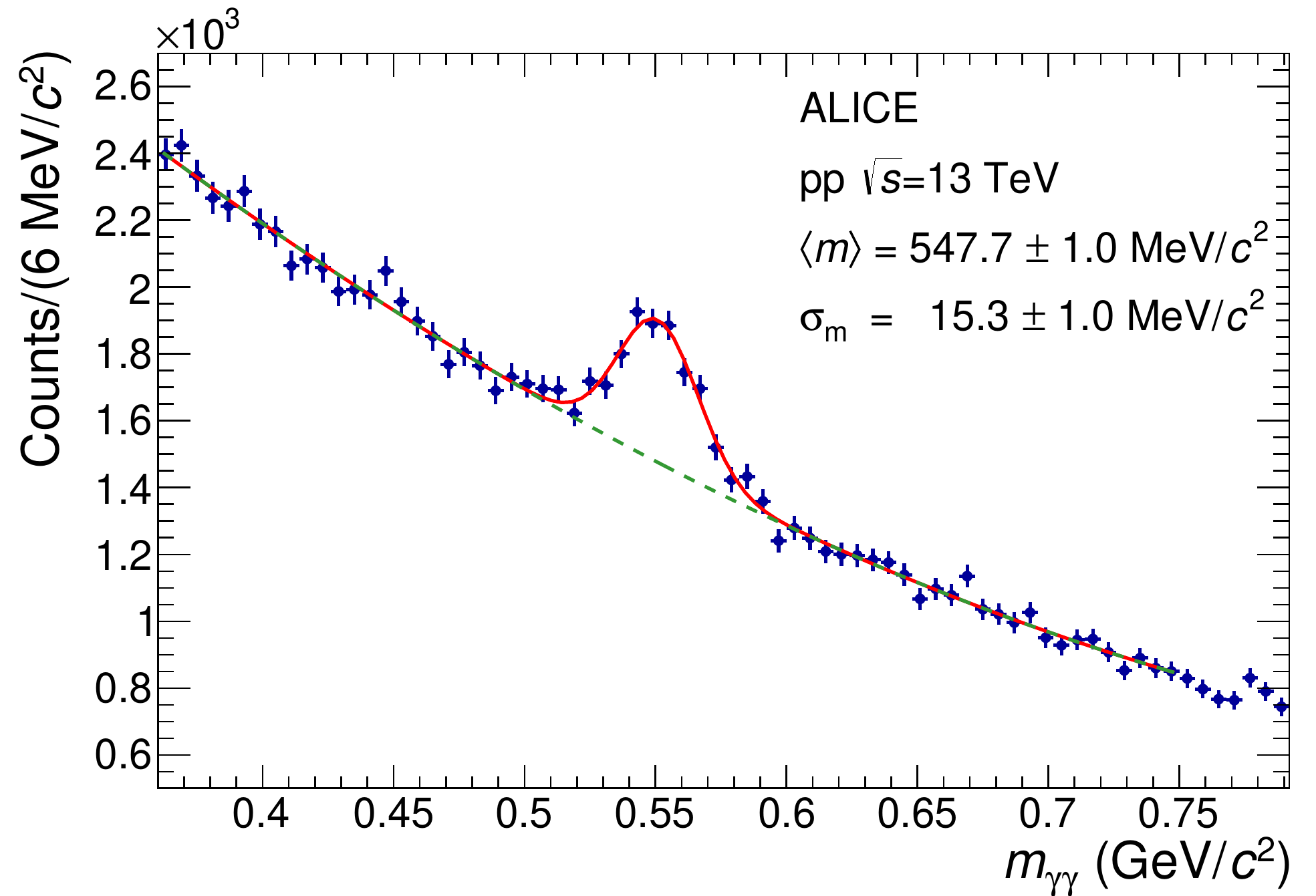}
  \caption{[Color online] Invariant mass distributions of cluster pairs for $\pT>1.7~\GeVc$ in
    the $\pi^0$ (left) and $\eta$ (right) mass region after calibration with per-channel $\pi^0$ peak
    equalization. For the $\pi^0$ data, the solid curve shows the fitting function using the
sum of the Crystal Ball and a polynomial function. For the $\eta$ data, the solid curve shows the fit function composed of a Gaussian and a polynomial function. The dashed lines represent the background contributions in both plots.}
  \label{fig:pi0PeakFinal}
\end{figure}

The invariant mass spectrum of cluster pairs, after applying all calibration corrections, is shown in Fig.~\ref{fig:pi0PeakFinal} in the region of the $\pi^0$ (left) and $\eta$-meson (right) peaks. All four PHOS modules were considered. It reveals a much narrower $\pi^0$ peak and better
signal-to-background ratio compared to the pre-calibrated result shown in Fig. \ref{fig:pi0PeakAPDgains}. 
The improved calibration allows to resolve details of the shape of the $\pi^0$ peak, therefore the mass distribution is fitted with a sum of a Crystal Ball function \cite{CrystalBall:1980} for the peak description, and a polynomial of the second order for the combinatorial background. For the $\eta$ meson a sum of Gaussian and second order polynomial is used.
Both the $\pi^0$ and $\eta$ meson peak positions are consistent with their PDG values of $m_{\pi^0}=134.98$~MeV$/c^2$ and $m_\eta=547.9$~MeV$/c^2$ within the statistical uncertainties shown in Fig.~\ref{fig:pi0PeakFinal}. 
The agreement of the $\eta$ peak position with the PDG values provides a cross-check of the correctness of the description of the PHOS alignment in the ALICE setup and therefore, of the absolute energy calibration.

\begin{figure}[h]
\centering
\includegraphics[width=0.48\linewidth]{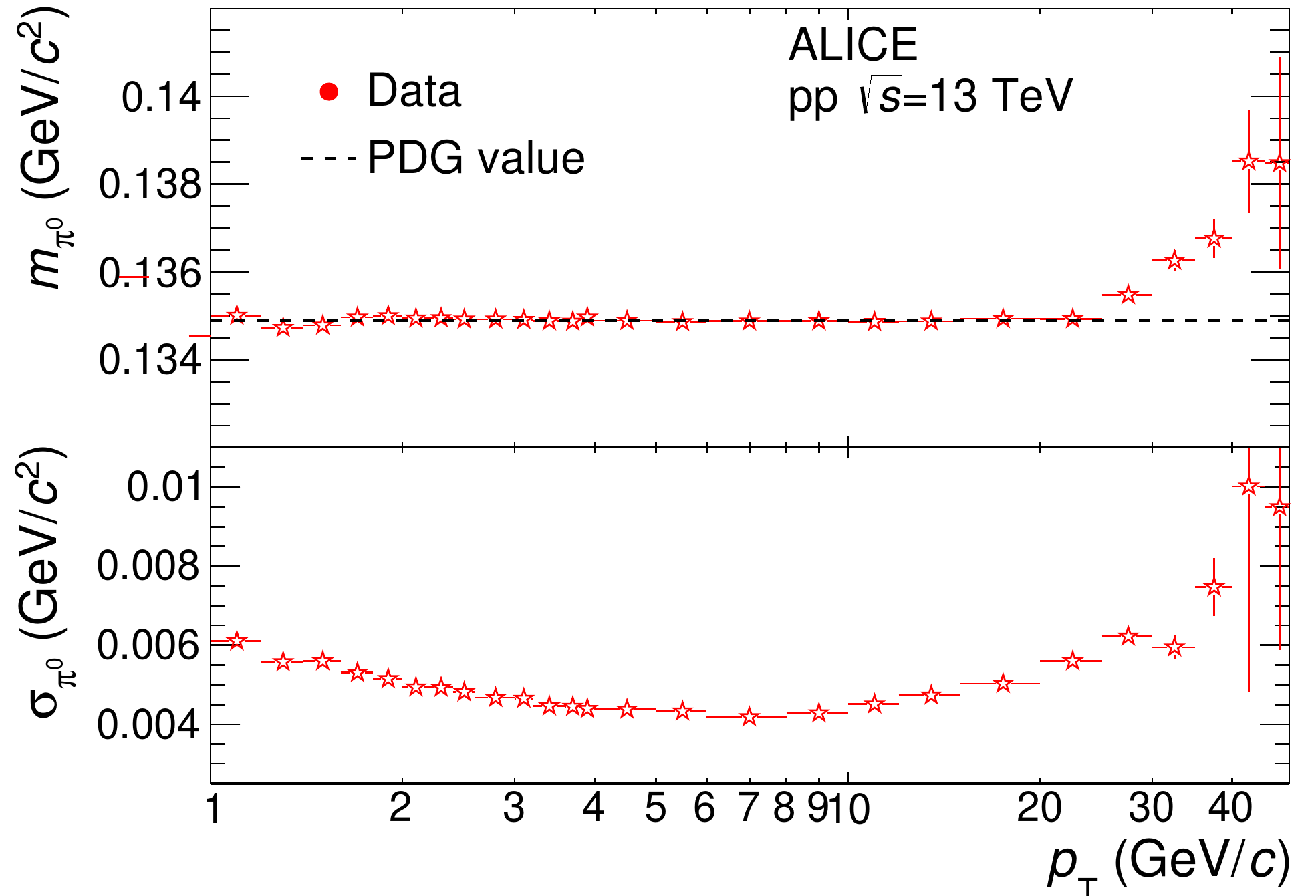}
\hfill
\includegraphics[width=0.48\linewidth]{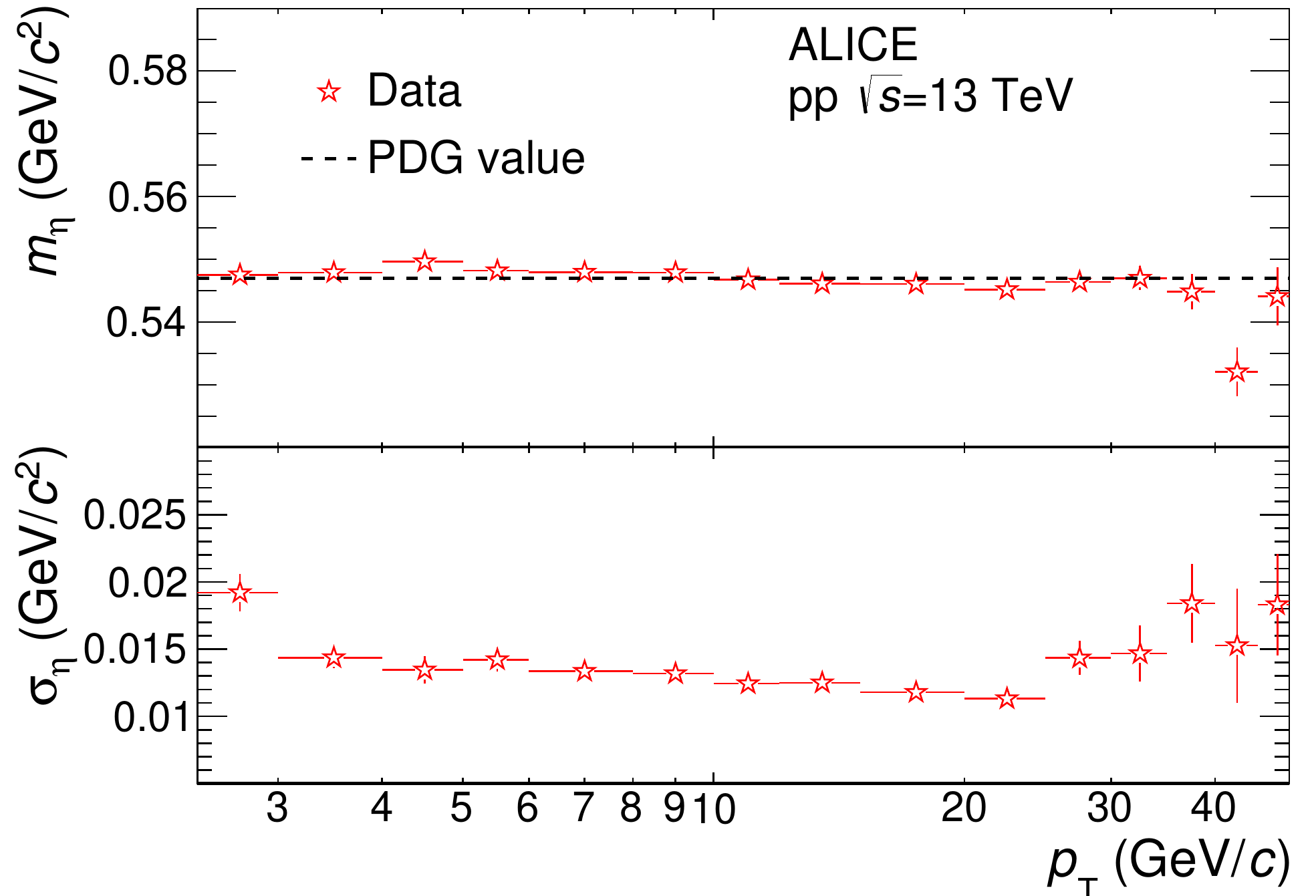}
\caption{
\label{fig:MassPi0} [Color online]
Peak position and width for $\pi^0$ (left) and $\eta$ mesons (right) as a function of transverse momentum. Vertical error bars represent fit uncertainties. }
\end{figure}

Figure \ref{fig:MassPi0} shows the peak positions and peak widths of the $\pi^0$ and $\eta$ mesons as a function of transverse momentum. The width of the $\pi^0$ peak reaches a minimum value $\sigma\approx4$~MeV$/c^2$ at $\pT=3-8~\GeVc$.
The reconstructed mass remains approximately constant up to $\pT\sim 25~\GeVc$, and  increases with $\pT$ afterwards. 
This is due to a considerable fraction of overlapping cluster pairs. The reconstruction software has a bias towards clusters that are better separated due to fluctuations in the energy deposition, thus increasing the extracted pion mass. This effect is not corrected for, instead MC simulations are used to account for it in the efficiency calculations. 
In the case of the $\eta$ meson, the peak position is stable since the influence of the overlap in this case only appear above $\pT\sim 80~\GeVc$. 

\section{Conclusions}

In this paper all
the steps of the calibration of the ALICE electromagnetic calorimeter PHOS from a completely uncalibrated state to the
final set of calibration parameters are presented. The results are equivalent to Monte Carlo simulations with an ideally calibrated detector. Pre-calibration, with the equalization of the photodetector gains,
is provided by the use of the monitoring system with light-emitting
diodes. This preliminary calibration serves as a starting point for the energy
calibration based on adjusting the reconstructed $\pi^0$ mass from
data collected in high-luminosity proton-proton
collisions. The calibration coefficients averaged over a large period of data taking are obtained with this 
relative calibration procedure. 

The absolute energy scale is verified by analyzing pp data with
electron tracks reconstructed in the ALICE central tracking system and
matched with PHOS clusters. An accurate correction of the PHOS geometrical alignment
in the radial direction, also achieved 
using electron tracks, is necessary for the absolute energy calibration. Further
refining of the calibration is performed by correcting the PHOS response
for energy nonlinearity effects. Finally, the calibration is corrected
for time variations in performance due to changes in running
conditions and power dissipation in the front-end electronics of the
detector. 

The resulting time-dependent calibration parameters of the
PHOS spectrometer ensure a stable response and the best possible
resolution of the detector over a large time span. 
After applying all calibration steps in the reconstruction of pp collision data at $\sqrt{s}=13$ TeV, 
the $\pi^0$ and $\eta$ meson peak positions are close to their PDG mass values over a wide $\pT$ range. The achieved mass resolution is $\sigma_m^{\pi^0} = 4.56 \pm 0.03$~MeV/$c^2$ and $\sigma_m^{\eta} = 15.3 \pm 1.0$~MeV/$c^2$ (for $\pT > 1.7$~GeV/$c$).

\newenvironment{acknowledgement}{\relax}{\relax}
\begin{acknowledgement}
\section*{Acknowledgements}

The ALICE Collaboration would like to thank all its engineers and technicians for their invaluable contributions to the construction of the experiment and the CERN accelerator teams for the outstanding performance of the LHC complex.
The ALICE Collaboration gratefully acknowledges the resources and support provided by all Grid centres and the Worldwide LHC Computing Grid (WLCG) collaboration.
The ALICE Collaboration acknowledges the following funding agencies for their support in building and running the ALICE detector:
A. I. Alikhanyan National Science Laboratory (Yerevan Physics Institute) Foundation (ANSL), State Committee of Science and World Federation of Scientists (WFS), Armenia;
Austrian Academy of Sciences, Austrian Science Fund \{FWF\}: [M 2467-N36] and Nationalstiftung f\"{u}r Forschung, Technologie und Entwicklung, Austria;
Ministry of Communications and High Technologies, National Nuclear Research Center, Azerbaijan;
Conselho Nacional de Desenvolvimento Cient\'{\i}fico e Tecnol\'{o}gico (CNPq), Universidade Federal do Rio Grande do Sul (UFRGS), Financiadora de Estudos e Projetos (Finep) and Funda\c{c}\~{a}o de Amparo \`{a} Pesquisa do Estado de S\~{a}o Paulo (FAPESP), Brazil;
Ministry of Science \& Technology of China (MSTC), National Natural Science Foundation of China (NSFC) and Ministry of Education of China (MOEC) , China;
Croatian Science Foundation and Ministry of Science and Education, Croatia;
Centro de Aplicaciones Tecnol\'{o}gicas y Desarrollo Nuclear (CEADEN), Cubaenerg\'{\i}a, Cuba;
Ministry of Education, Youth and Sports of the Czech Republic, Czech Republic;
The Danish Council for Independent Research | Natural Sciences, the Carlsberg Foundation and Danish National Research Foundation (DNRF), Denmark;
Helsinki Institute of Physics (HIP), Finland;
Commissariat \`{a} l'Energie Atomique (CEA), Institut National de Physique Nucl\'{e}aire et de Physique des Particules (IN2P3) and Centre National de la Recherche Scientifique (CNRS) and Rl\'{e}gion des  Pays de la Loire, France;
Bundesministerium f\"{u}r Bildung, Wissenschaft, Forschung und Technologie (BMBF) and GSI Helmholtzzentrum f\"{u}r Schwerionenforschung GmbH, Germany;
General Secretariat for Research and Technology, Ministry of Education, Research and Religions, Greece;
National Research, Development and Innovation Office, Hungary;
Department of Atomic Energy Government of India (DAE), Department of Science and Technology, Government of India (DST), University Grants Commission, Government of India (UGC) and Council of Scientific and Industrial Research (CSIR), India;
Indonesian Institute of Science, Indonesia;
Centro Fermi - Museo Storico della Fisica e Centro Studi e Ricerche Enrico Fermi and Istituto Nazionale di Fisica Nucleare (INFN), Italy;
Institute for Innovative Science and Technology , Nagasaki Institute of Applied Science (IIST), Japan Society for the Promotion of Science (JSPS) KAKENHI and Japanese Ministry of Education, Culture, Sports, Science and Technology (MEXT), Japan;
Consejo Nacional de Ciencia (CONACYT) y Tecnolog\'{i}a, through Fondo de Cooperaci\'{o}n Internacional en Ciencia y Tecnolog\'{i}a (FONCICYT) and Direcci\'{o}n General de Asuntos del Personal Academico (DGAPA), Mexico;
Nederlandse Organisatie voor Wetenschappelijk Onderzoek (NWO), Netherlands;
The Research Council of Norway, Norway;
Commission on Science and Technology for Sustainable Development in the South (COMSATS), Pakistan;
Pontificia Universidad Cat\'{o}lica del Per\'{u}, Peru;
Ministry of Science and Higher Education and National Science Centre, Poland;
Korea Institute of Science and Technology Information and National Research Foundation of Korea (NRF), Republic of Korea;
Ministry of Education and Scientific Research, Institute of Atomic Physics and Ministry of Research and Innovation and Institute of Atomic Physics, Romania;
Joint Institute for Nuclear Research (JINR), Ministry of Education and Science of the Russian Federation, National Research Centre Kurchatov Institute, Russian Science Foundation and Russian Foundation for Basic Research, Russia;
Ministry of Education, Science, Research and Sport of the Slovak Republic, Slovakia;
National Research Foundation of South Africa, South Africa;
Swedish Research Council (VR) and Knut \& Alice Wallenberg Foundation (KAW), Sweden;
European Organization for Nuclear Research, Switzerland;
National Science and Technology Development Agency (NSDTA), Suranaree University of Technology (SUT) and Office of the Higher Education Commission under NRU project of Thailand, Thailand;
Turkish Atomic Energy Agency (TAEK), Turkey;
National Academy of  Sciences of Ukraine, Ukraine;
Science and Technology Facilities Council (STFC), United Kingdom;
National Science Foundation of the United States of America (NSF) and United States Department of Energy, Office of Nuclear Physics (DOE NP), United States of America.    
\end{acknowledgement}

\bibliographystyle{utphys} 
\bibliography{CalibrationPHOS}


\newpage
\appendix
\section{The ALICE Collaboration}
\label{app:collab}

\begingroup
\small
\begin{flushleft}
S.~Acharya\Irefn{org140}\And 
F.T.-.~Acosta\Irefn{org20}\And 
D.~Adamov\'{a}\Irefn{org93}\And 
S.P.~Adhya\Irefn{org140}\And 
A.~Adler\Irefn{org74}\And 
J.~Adolfsson\Irefn{org80}\And 
M.M.~Aggarwal\Irefn{org98}\And 
G.~Aglieri Rinella\Irefn{org34}\And 
M.~Agnello\Irefn{org31}\And 
Z.~Ahammed\Irefn{org140}\And 
S.~Ahmad\Irefn{org17}
% ...
\And
\renewcommand\labelenumi{\textsuperscript{\theenumi}~}

\section*{Affiliation notes}
\renewcommand\theenumi{\roman{enumi}}
\begin{Authlist}
\item \Adef{org*}Deceased
\item \Adef{orgI}Dipartimento DET del Politecnico di Torino, Turin, Italy
\item \Adef{orgII}M.V. Lomonosov Moscow State University, D.V. Skobeltsyn Institute of Nuclear, Physics, Moscow, Russia
\item \Adef{orgIII}Department of Applied Physics, Aligarh Muslim University, Aligarh, India
\item \Adef{orgIV}Institute of Theoretical Physics, University of Wroclaw, Poland
\end{Authlist}

\section*{Collaboration Institutes}
\renewcommand\theenumi{\arabic{enumi}~}
\begin{Authlist}
\item \Idef{org1}A.I. Alikhanyan National Science Laboratory (Yerevan Physics Institute) Foundation, Yerevan, Armenia
\item \Idef{org2}Bogolyubov Institute for Theoretical Physics, National Academy of Sciences of Ukraine, Kiev, Ukraine
\item \Idef{org3}Bose Institute, Department of Physics  and Centre for Astroparticle Physics and Space Science (CAPSS), Kolkata, India
\item \Idef{org4}Budker Institute for Nuclear Physics, Novosibirsk, Russia
\item \Idef{org5}California Polytechnic State University, San Luis Obispo, California, United States
\item \Idef{org6}Central China Normal University, Wuhan, China
\item \Idef{org7}Centre de Calcul de l'IN2P3, Villeurbanne, Lyon, France
\item \Idef{org8}Centro de Aplicaciones Tecnol\'{o}gicas y Desarrollo Nuclear (CEADEN), Havana, Cuba
\item \Idef{org9}Centro de Investigaci\'{o}n y de Estudios Avanzados (CINVESTAV), Mexico City and M\'{e}rida, Mexico
\item \Idef{org10}Centro Fermi - Museo Storico della Fisica e Centro Studi e Ricerche ``Enrico Fermi', Rome, Italy
\item \Idef{org11}Chicago State University, Chicago, Illinois, United States
\item \Idef{org12}China Institute of Atomic Energy, Beijing, China
\item \Idef{org13}Chonbuk National University, Jeonju, Republic of Korea
\item \Idef{org14}Comenius University Bratislava, Faculty of Mathematics, Physics and Informatics, Bratislava, Slovakia
\item \Idef{org15}COMSATS Institute of Information Technology (CIIT), Islamabad, Pakistan
\item \Idef{org16}Creighton University, Omaha, Nebraska, United States
\item \Idef{org17}Department of Physics, Aligarh Muslim University, Aligarh, India
\item \Idef{org18}Department of Physics, Pusan National University, Pusan, Republic of Korea
\item \Idef{org19}Department of Physics, Sejong University, Seoul, Republic of Korea
\item \Idef{org20}Department of Physics, University of California, Berkeley, California, United States
\item \Idef{org21}Department of Physics, University of Oslo, Oslo, Norway
\item \Idef{org22}Department of Physics and Technology, University of Bergen, Bergen, Norway
\item \Idef{org23}Dipartimento di Fisica dell'Universit\`{a} 'La Sapienza' and Sezione INFN, Rome, Italy
\item \Idef{org24}Dipartimento di Fisica dell'Universit\`{a} and Sezione INFN, Cagliari, Italy
\item \Idef{org25}Dipartimento di Fisica dell'Universit\`{a} and Sezione INFN, Trieste, Italy
\item \Idef{org26}Dipartimento di Fisica dell'Universit\`{a} and Sezione INFN, Turin, Italy
\item \Idef{org27}Dipartimento di Fisica e Astronomia dell'Universit\`{a} and Sezione INFN, Bologna, Italy
\item \Idef{org28}Dipartimento di Fisica e Astronomia dell'Universit\`{a} and Sezione INFN, Catania, Italy
\item \Idef{org29}Dipartimento di Fisica e Astronomia dell'Universit\`{a} and Sezione INFN, Padova, Italy
\item \Idef{org30}Dipartimento di Fisica `E.R.~Caianiello' dell'Universit\`{a} and Gruppo Collegato INFN, Salerno, Italy
\item \Idef{org31}Dipartimento DISAT del Politecnico and Sezione INFN, Turin, Italy
\item \Idef{org32}Dipartimento di Scienze e Innovazione Tecnologica dell'Universit\`{a} del Piemonte Orientale and INFN Sezione di Torino, Alessandria, Italy
\item \Idef{org33}Dipartimento Interateneo di Fisica `M.~Merlin' and Sezione INFN, Bari, Italy
\item \Idef{org34}European Organization for Nuclear Research (CERN), Geneva, Switzerland
\item \Idef{org35}Faculty of Electrical Engineering, Mechanical Engineering and Naval Architecture, University of Split, Split, Croatia
\item \Idef{org36}Faculty of Engineering and Science, Western Norway University of Applied Sciences, Bergen, Norway
\item \Idef{org37}Faculty of Nuclear Sciences and Physical Engineering, Czech Technical University in Prague, Prague, Czech Republic
\item \Idef{org38}Faculty of Science, P.J.~\v{S}af\'{a}rik University, Ko\v{s}ice, Slovakia
\item \Idef{org39}Frankfurt Institute for Advanced Studies, Johann Wolfgang Goethe-Universit\"{a}t Frankfurt, Frankfurt, Germany
\item \Idef{org40}Gangneung-Wonju National University, Gangneung, Republic of Korea
\item \Idef{org41}Gauhati University, Department of Physics, Guwahati, India
\item \Idef{org42}Helmholtz-Institut f\"{u}r Strahlen- und Kernphysik, Rheinische Friedrich-Wilhelms-Universit\"{a}t Bonn, Bonn, Germany
\item \Idef{org43}Helsinki Institute of Physics (HIP), Helsinki, Finland
\item \Idef{org44}High Energy Physics Group,  Universidad Aut\'{o}noma de Puebla, Puebla, Mexico
\item \Idef{org45}Hiroshima University, Hiroshima, Japan
\item \Idef{org46}Hochschule Worms, Zentrum  f\"{u}r Technologietransfer und Telekommunikation (ZTT), Worms, Germany
\item \Idef{org47}Horia Hulubei National Institute of Physics and Nuclear Engineering, Bucharest, Romania
\item \Idef{org48}Indian Institute of Technology Bombay (IIT), Mumbai, India
\item \Idef{org49}Indian Institute of Technology Indore, Indore, India
\item \Idef{org50}Indonesian Institute of Sciences, Jakarta, Indonesia
\item \Idef{org51}INFN, Laboratori Nazionali di Frascati, Frascati, Italy
\item \Idef{org52}INFN, Sezione di Bari, Bari, Italy
\item \Idef{org53}INFN, Sezione di Bologna, Bologna, Italy
\item \Idef{org54}INFN, Sezione di Cagliari, Cagliari, Italy
\item \Idef{org55}INFN, Sezione di Catania, Catania, Italy
\item \Idef{org56}INFN, Sezione di Padova, Padova, Italy
\item \Idef{org57}INFN, Sezione di Roma, Rome, Italy
\item \Idef{org58}INFN, Sezione di Torino, Turin, Italy
\item \Idef{org59}INFN, Sezione di Trieste, Trieste, Italy
\item \Idef{org60}Inha University, Incheon, Republic of Korea
\item \Idef{org61}Institut de Physique Nucl\'{e}aire d'Orsay (IPNO), Institut National de Physique Nucl\'{e}aire et de Physique des Particules (IN2P3/CNRS), Universit\'{e} de Paris-Sud, Universit\'{e} Paris-Saclay, Orsay, France
\item \Idef{org62}Institute for Nuclear Research, Academy of Sciences, Moscow, Russia
\item \Idef{org63}Institute for Subatomic Physics, Utrecht University/Nikhef, Utrecht, Netherlands
\item \Idef{org64}Institute for Theoretical and Experimental Physics, Moscow, Russia
\item \Idef{org65}Institute of Experimental Physics, Slovak Academy of Sciences, Ko\v{s}ice, Slovakia
\item \Idef{org66}Institute of Physics, Homi Bhabha National Institute, Bhubaneswar, India
\item \Idef{org67}Institute of Physics of the Czech Academy of Sciences, Prague, Czech Republic
\item \Idef{org68}Institute of Space Science (ISS), Bucharest, Romania
\item \Idef{org69}Institut f\"{u}r Kernphysik, Johann Wolfgang Goethe-Universit\"{a}t Frankfurt, Frankfurt, Germany
\item \Idef{org70}Instituto de Ciencias Nucleares, Universidad Nacional Aut\'{o}noma de M\'{e}xico, Mexico City, Mexico
\item \Idef{org71}Instituto de F\'{i}sica, Universidade Federal do Rio Grande do Sul (UFRGS), Porto Alegre, Brazil
\item \Idef{org72}Instituto de F\'{\i}sica, Universidad Nacional Aut\'{o}noma de M\'{e}xico, Mexico City, Mexico
\item \Idef{org73}iThemba LABS, National Research Foundation, Somerset West, South Africa
\item \Idef{org74}Johann-Wolfgang-Goethe Universit\"{a}t Frankfurt Institut f\"{u}r Informatik, Fachbereich Informatik und Mathematik, Frankfurt, Germany
\item \Idef{org75}Joint Institute for Nuclear Research (JINR), Dubna, Russia
\item \Idef{org76}Korea Institute of Science and Technology Information, Daejeon, Republic of Korea
\item \Idef{org77}KTO Karatay University, Konya, Turkey
\item \Idef{org78}Laboratoire de Physique Subatomique et de Cosmologie, Universit\'{e} Grenoble-Alpes, CNRS-IN2P3, Grenoble, France
\item \Idef{org79}Lawrence Berkeley National Laboratory, Berkeley, California, United States
\item \Idef{org80}Lund University Department of Physics, Division of Particle Physics, Lund, Sweden
\item \Idef{org81}Nagasaki Institute of Applied Science, Nagasaki, Japan
\item \Idef{org82}Nara Women{'}s University (NWU), Nara, Japan
\item \Idef{org83}National and Kapodistrian University of Athens, School of Science, Department of Physics , Athens, Greece
\item \Idef{org84}National Centre for Nuclear Research, Warsaw, Poland
\item \Idef{org85}National Institute of Science Education and Research, Homi Bhabha National Institute, Jatni, India
\item \Idef{org86}National Nuclear Research Center, Baku, Azerbaijan
\item \Idef{org87}National Research Centre Kurchatov Institute, Moscow, Russia
\item \Idef{org88}Niels Bohr Institute, University of Copenhagen, Copenhagen, Denmark
\item \Idef{org89}Nikhef, National institute for subatomic physics, Amsterdam, Netherlands
\item \Idef{org90}NRC Kurchatov Institute IHEP, Protvino, Russia
\item \Idef{org91}NRNU Moscow Engineering Physics Institute, Moscow, Russia
\item \Idef{org92}Nuclear Physics Group, STFC Daresbury Laboratory, Daresbury, United Kingdom
\item \Idef{org93}Nuclear Physics Institute of the Czech Academy of Sciences, \v{R}e\v{z} u Prahy, Czech Republic
\item \Idef{org94}Oak Ridge National Laboratory, Oak Ridge, Tennessee, United States
\item \Idef{org95}Ohio State University, Columbus, Ohio, United States
\item \Idef{org96}Petersburg Nuclear Physics Institute, Gatchina, Russia
\item \Idef{org97}Physics department, Faculty of science, University of Zagreb, Zagreb, Croatia
\item \Idef{org98}Physics Department, Panjab University, Chandigarh, India
\item \Idef{org99}Physics Department, University of Jammu, Jammu, India
\item \Idef{org100}Physics Department, University of Rajasthan, Jaipur, India
\item \Idef{org101}Physikalisches Institut, Eberhard-Karls-Universit\"{a}t T\"{u}bingen, T\"{u}bingen, Germany
\item \Idef{org102}Physikalisches Institut, Ruprecht-Karls-Universit\"{a}t Heidelberg, Heidelberg, Germany
\item \Idef{org103}Physik Department, Technische Universit\"{a}t M\"{u}nchen, Munich, Germany
\item \Idef{org104}Research Division and ExtreMe Matter Institute EMMI, GSI Helmholtzzentrum f\"ur Schwerionenforschung GmbH, Darmstadt, Germany
\item \Idef{org105}Rudjer Bo\v{s}kovi\'{c} Institute, Zagreb, Croatia
\item \Idef{org106}Russian Federal Nuclear Center (VNIIEF), Sarov, Russia
\item \Idef{org107}Saha Institute of Nuclear Physics, Homi Bhabha National Institute, Kolkata, India
\item \Idef{org108}School of Physics and Astronomy, University of Birmingham, Birmingham, United Kingdom
\item \Idef{org109}Secci\'{o}n F\'{\i}sica, Departamento de Ciencias, Pontificia Universidad Cat\'{o}lica del Per\'{u}, Lima, Peru
\item \Idef{org110}Shanghai Institute of Applied Physics, Shanghai, China
\item \Idef{org111}St. Petersburg State University, St. Petersburg, Russia
\item \Idef{org112}Stefan Meyer Institut f\"{u}r Subatomare Physik (SMI), Vienna, Austria
\item \Idef{org113}SUBATECH, IMT Atlantique, Universit\'{e} de Nantes, CNRS-IN2P3, Nantes, France
\item \Idef{org114}Suranaree University of Technology, Nakhon Ratchasima, Thailand
\item \Idef{org115}Technical University of Ko\v{s}ice, Ko\v{s}ice, Slovakia
\item \Idef{org116}Technische Universit\"{a}t M\"{u}nchen, Excellence Cluster 'Universe', Munich, Germany
\item \Idef{org117}The Henryk Niewodniczanski Institute of Nuclear Physics, Polish Academy of Sciences, Cracow, Poland
\item \Idef{org118}The University of Texas at Austin, Austin, Texas, United States
\item \Idef{org119}Universidad Aut\'{o}noma de Sinaloa, Culiac\'{a}n, Mexico
\item \Idef{org120}Universidade de S\~{a}o Paulo (USP), S\~{a}o Paulo, Brazil
\item \Idef{org121}Universidade Estadual de Campinas (UNICAMP), Campinas, Brazil
\item \Idef{org122}Universidade Federal do ABC, Santo Andre, Brazil
\item \Idef{org123}University College of Southeast Norway, Tonsberg, Norway
\item \Idef{org124}University of Cape Town, Cape Town, South Africa
\item \Idef{org125}University of Houston, Houston, Texas, United States
\item \Idef{org126}University of Jyv\"{a}skyl\"{a}, Jyv\"{a}skyl\"{a}, Finland
\item \Idef{org127}University of Liverpool, Liverpool, United Kingdom
\item \Idef{org128}University of Science and Techonology of China, Hefei, China
\item \Idef{org129}University of Tennessee, Knoxville, Tennessee, United States
\item \Idef{org130}University of the Witwatersrand, Johannesburg, South Africa
\item \Idef{org131}University of Tokyo, Tokyo, Japan
\item \Idef{org132}University of Tsukuba, Tsukuba, Japan
\item \Idef{org133}Universit\'{e} Clermont Auvergne, CNRS/IN2P3, LPC, Clermont-Ferrand, France
\item \Idef{org134}Universit\'{e} de Lyon, Universit\'{e} Lyon 1, CNRS/IN2P3, IPN-Lyon, Villeurbanne, Lyon, France
\item \Idef{org135}Universit\'{e} de Strasbourg, CNRS, IPHC UMR 7178, F-67000 Strasbourg, France, Strasbourg, France
\item \Idef{org136} Universit\'{e} Paris-Saclay Centre d¿\'Etudes de Saclay (CEA), IRFU, Department de Physique Nucl\'{e}aire (DPhN), Saclay, France
\item \Idef{org137}Universit\`{a} degli Studi di Foggia, Foggia, Italy
\item \Idef{org138}Universit\`{a} degli Studi di Pavia, Pavia, Italy
\item \Idef{org139}Universit\`{a} di Brescia, Brescia, Italy
\item \Idef{org140}Variable Energy Cyclotron Centre, Homi Bhabha National Institute, Kolkata, India
\item \Idef{org141}Warsaw University of Technology, Warsaw, Poland
\item \Idef{org142}Wayne State University, Detroit, Michigan, United States
\item \Idef{org143}Westf\"{a}lische Wilhelms-Universit\"{a}t M\"{u}nster, Institut f\"{u}r Kernphysik, M\"{u}nster, Germany
\item \Idef{org144}Wigner Research Centre for Physics, Hungarian Academy of Sciences, Budapest, Hungary
\item \Idef{org145}Yale University, New Haven, Connecticut, United States
\item \Idef{org146}Yonsei University, Seoul, Republic of Korea
\end{Authlist}
\endgroup
\end{document}